\documentstyle[12pt,graphicx,amssymb]{article} 
\title{\large
\vspace{-3cm}
                \hspace*{7cm}DESY 79-048 \\ 
                \hspace*{7cm}ECFA-97-182 \\
                 \hspace*{6cm}Budker INP 97-57 \\
\vspace{2cm}
\LARGE\bf An Interaction Region  for Gamma-Gamma and Gamma-Electron
Collisions  at TESLA/SBLC}

\author{R.~Brinkmann$^1$, I.~Ginzburg$^2$, N.~Holtkamp$^1$,
G.~Jikia$^3$, \\ O.~Napoly$^4$, E.~Saldin$^5$, E.~Schneidmiller$^5$,
V.~Serbo$^6$, \\ G.~Silvestrov$^7$,  
V.~Telnov$^7$\thanks{corresponding author:
telnov@inp.nsk.su} , A.~Undrus$^7$,  M.~Yurkov$^8$} 
\date{}
\begin{document}
\newcommand{\EP}{\mbox{e$^+$}}
\newcommand{\EM}{\mbox{e$^-$}}
\newcommand{\EPEM}{\mbox{e$^+$e$^-$}}
\newcommand{\EMEM}{\mbox{e$^-$e$^-$}}
\newcommand{\EE}{\mbox{ee}}
\newcommand{\GG}{\mbox{$\gamma\gamma$}}
\newcommand{\GP}{\mbox{$\gamma$e$^+$}}
\newcommand{\GE}{\mbox{$\gamma$e}}
\newcommand{\LGE}{\mbox{$L_{\GE}$}}
\newcommand{\LGG}{\mbox{$L_{\GG}$}}
\newcommand{\LEE}{\mbox{$L_{\EE}$}}
\newcommand{\TEV}{\mbox{TeV}}
\newcommand{\GEV}{\mbox{GeV}}
\newcommand{\EV}{\mbox{eV}}
\newcommand{\CM}{\mbox{cm}}
\newcommand{\MM}{\mbox{mm}}
\newcommand{\NM}{\mbox{nm}}
\newcommand{\MKM}{\mbox{$\mu$m}}
\newcommand{\EXN}{\mbox{$\epsilon_{nx}$}}
\newcommand{\EYN}{\mbox{$\epsilon_{ny}$}}
\newcommand{\SEC}{\mbox{s}}
\newcommand{\CMS}{\mbox{cm$^{-2}$s$^{-1}$}}
\newcommand{\MRAD}{\mbox{mrad}}
\newcommand{\IND}{\hspace*{\parindent}}
\newcommand{\beq}{\begin{equation}}
\newcommand{\eeq}{\end{equation}}
\newcommand{\beqn}{\begin{eqnarray}}
\newcommand{\eeqn}{\end{eqnarray}}
\newcommand{\dst}{\displaystyle}
\newcommand{\bm}{\boldmath}
\newcommand{\fr}[2]{\frac{{\dst #1}}{{\dst #2}}}
\newcommand{\lum}[1]{{\rm luminosity} $ #1 $ cm$^{-2}$ s$^{-1}\,$}
\newcommand{\intlum}[1]{{\rm annual luminosity} $ #1$  fb$^{-1}\,$}
\newcommand{\mw}{\mbox{$M_W\,$}}
\newcommand{\mww}{\mbox{$M_W^2\,$}}
\newcommand{\mh}{\mbox{$M_H\,$}}
\newcommand{\mhh}{\mbox{$M_H^2\,$}}
\newcommand{\mz}{\mbox{$M_Z\,$}}
\newcommand{\mzz}{\mbox{$M_Z^2\,$}}
\newcommand{\sigmaw}{\mbox{$\sigma_W\,$}}
\newcommand{\sw}{\mbox{$\sin\Theta_W\,$}}
\newcommand{\sww}{\mbox{$\sin^2\Theta_W\,$}}
\newcommand{\cw}{\mbox{$\cos\Theta_W\,$}}
\newcommand{\cww}{\mbox{$\cos^2\Theta_W\,$}}
\newcommand{\ptr}{\mbox{$p_{\bot}\,$}}
\newcommand{\ptrs}{\mbox{$p_{\bot}^2\,$}}
\newcommand{\lgam}{\mbox{$\lambda_{\gamma}$}}
\newcommand{\lga}[1]{\mbox{$\lambda_{#1}$}}
\newcommand{\lggam}[2]{\mbox{$\lambda_{#1}\lambda_{#2}$}}
\newcommand{\lgg}{\lambda_1\lambda_2}
\newcommand{\lel}{\mbox{$\lambda_e$}}
\newcommand{\epe}{\mbox{$e^+e^-\,$}}
\newcommand{\ggam}{\mbox{$\gamma\gamma\,$}}
\newcommand{\egam}{\mbox{$\gamma e\,$}}
\newcommand{\gewnu}{\mbox{$\gamma e\to \textrm{\rm W}\nu\,$}}
\newcommand{\eeww}{\mbox{$e^+e^-\to \textrm{\rm W}^+\textrm{\rm W}^-\,$}}
\newcommand{\ggww}{\mbox{$\gamma\gamma\to \textrm{\rm W}^+\textrm{\rm W}^-\,$}}
\newcommand{\ggzz}{\mbox{$\gamma\gamma\to \textrm{ZZ}\,$}}
\newcommand{\geeww}{\mbox{$\gamma e\to  \textrm{\rm W}^+\textrm{\rm W}^- e\,$}}
\newcommand{\ggh}{\mbox{$\gamma\gamma\to hadrons$}}
\newcommand{\pair}[1]{\mbox{$#1 \bar{#1}$}}
\newcommand{\ZZ}{\mbox{ZZ}}
\newcommand{\WW}{\mbox{WW}}
\newcommand{\Z}{\mbox{Z}}
\newcommand{\W}{\mbox{W}}
\maketitle
\vspace*{3cm} 
----------------------------- \\ 
1) DESY, Germany \\ 
2) Institute of Mathematics, Novosibirsk, Russia\\  
3) University of Freiburg, Germany and IHEP, Protvino, Russia \\
4) CEA, Saclay, France\\  
5) Automatic System Corporation, Samara, Russia\\  
6) Novosibirsk State University, Novosibirsk, Russia \\ 
7) Institute of Nuclear Physics, Novosibirsk, Russia\\  
8) JINR, Dubna, Russia
\newpage
\begin{abstract}

 Linear colliders offer an unique opportunities to study $\gamma\gamma$
and $\gamma$e interactions. Using the laser backscattering method one
can obtain $\gamma\gamma$, $\gamma$e colliding beams with an energy
and luminosity comparable to that in e$^+$e$^-$ collisions.  This work
is  part of the Conceptual Design of TESLA/SBLC linear colliders
describing  a second interaction region for $\gamma\gamma$ and
$\gamma$e collisions. We consider here possible physics in high energy
$\gamma\gamma$, $\gamma$e collisions, e$\to\gamma$ conversion,
requirements to lasers, collision schemes, attainable luminosities,
backgrounds, possible lasers, optics at the interaction region and
other associated problems.

\end{abstract}
\vspace*{1cm}

\hspace*{4cm}{\large\bf Contents} \\
{\bf 1.  Introduction. \\
2.  Physics at \GG,\GE\ collider}. \\
\hspace*{1cm}2.1$\;\;$Introduction. \\
\hspace*{1cm}2.2$\;\;$Higgs boson physics. \\
\hspace*{1cm}2.3$\;\;$Gauge boson physics. \\
\hspace*{1cm}2.4$\;\;$Physics of t-quark. \\
\hspace*{1cm}2.5$\;\;$Hadron physics and QCD. \\
{\bf 3.  Conversion region.} \\
\hspace*{1cm}3.1$\;\;$Choice of laser parameters, conversion efficiency.\\
\hspace*{1cm}3.2$\;\;$Low energy electrons after multiple Compton 
scatterings \\
{\bf 4.  Interaction region.}\\
\hspace*{1cm}4.1$\;\;$Collision schemes. \\
\hspace*{1cm}4.2$\;\;$Collision effects in \GG, \GE\ collisions. \\
\hspace*{1cm}4.3$\;\;$Simulation code. \\
\hspace*{1cm}4.4$\;\;$Parameters of electron beams at IP. \\
\hspace*{1cm}4.5$\;\;$Simulation results. \\
\hspace*{2cm}4.5.1$\;\;$Scheme without magnetic deflection; \\
\hspace*{2cm}4.5.2$\;\;$Scheme with magnetic deflection; \\
\hspace*{2cm}4.5.3$\;\;$Ultimate luminosities. \\
\hspace*{1cm}4.6$\;\;$Summary table of \GG, \GE\ luminosities. \\
\hspace*{1cm}4.7$\;\;$Monitoring and measurement of \GG, \GE\ luminosities.\\
\hspace*{1cm}4.8$\;\;$Sweeping magnet. \\
{\bf 5.  Backgrounds.  \\
6. Optics in the interaction region. \\
7. Lasers.  \\
}
\hspace*{2cm}7.1   Solid state lasers. \\
\hspace*{2cm}7.2 Free electron lasers. \\
\newpage

\section{Introduction}

Due to the synchrotron radiation problem in \EPEM\ storage ring the energy
region above LEP II can only be explored by linear colliders. There are several
project of linear colliders on the energy up to 2E $\sim$ 500--1000 GeV 
\cite{LOEW}.  

Linear colliders offer unique opportunities to study \GG, \GE\
interactions. Using the laser backscattering method one can obtain
\GG\ and \GE\ colliding beams with an energy and luminosity comparable
to that in \EPEM\ collisions. This can be done with relatively small
incremental cost. The expected physics in these collisions (see
section 2) is very rich and complementary to that in \EPEM\
collisions. Some characteristic examples are:

\begin{itemize}
\item a \GG\ collider provides unique opportunities to measure the
  two-photon decay width of the Higgs boson, and to search for
  relatively heavy Higgs states in the extended Higgs models such as
  MSSM;

\item a \GG\ collider is an outstanding $W$ factory, with a $WW$ pair
  production cross section by a factor of 10--20 larger than in
  \EPEM\, and with a potential of producing $10^6-10^7$ $W$'s per
  year, allowing a precision study of the anomalous gauge boson
  interactions;

\item a \GG, \GE\ collider is a remarkable tool for searching for new
charged particles, such as supersymmetric particles, leptoquarks,
excited states of electrons, etc., as in \GG, \GE\ collisions they are
produced with cross sections larger than in \EPEM\ collisions;

\item at a \GE\ collider charged supersymmetric particles with masses
  higher than the beam energy could be produced as well as the structure
of the photon could be measured with comparable presision to studies of the
proton at HERA..
\end{itemize}

In order to make this new field of particle physics accessible it
would be wise to have at the linear collider two interaction regions (IR): 
one for \EPEM\ collisions and the other for \GG, \GE\ collisions. 

This paper is a part of the TESLA/SBLC Conceptual Design \cite{TESLA}
developed at DESY in collaboration with many other institutes. Here we
will describe physics at photon colliders, expected \GG, \GE\
luminosities, and the design of the interaction region required for
\GG, \GE\ collisions.

\begin{figure}[!hbp]
\centering
\includegraphics[width=2in,angle=-90]{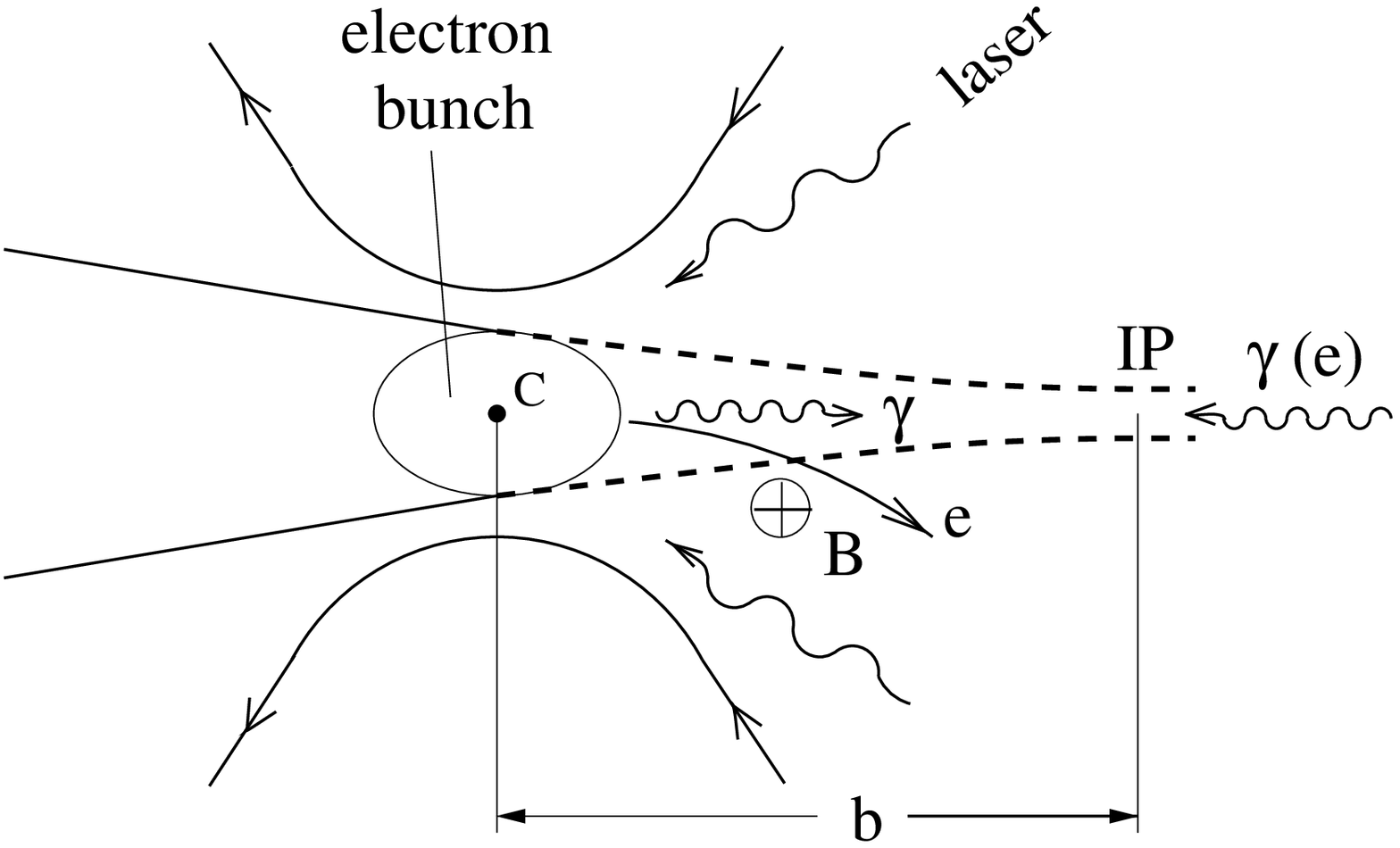}
\caption{Scheme of a \GG; \GE\ collider.}
\label{fig1}
\end{figure}
 
The general scheme of a photon collider is shown in Fig.~\ref{fig1}.  
Two electron beams after the final focus system are traveling
toward the interaction point (IP).  At a  distance of order 1 cm
upstream from the IP, referred to hereafter as the conversion point
(CP), the laser beam is focused and Compton backscattered by the
electrons, resulting in the high energy beam of photons. 
 With reasonable laser
parameters one can ``convert'' most of electrons to high energy
photons. The photon beam follows the original electron direction 
of motion with a small angular spread
of order $1/\gamma $, arriving at the IP in a tight focus, where it
collides with a similar opposing high energy photon beam or with an 
electron beam. The photon spot size at the IP may be almost
equal to that of electrons at IP and therefore the luminosity of
\GG, \GE\ collisions will be of the same
order as the ``geometric'' luminosity of basic $ee$ beams
(positrons are not necessary for a photon collider).

After multiple Compton scattering the electrons have a wide energy spread
$E=(0.02-1)E_0$, and either follow to the IP or are swept aside by a small
pulsed magnet with $B \sim 1$ T. With the deflecting magnet one can
get better quality of colliding beams, smaller background and 
disruption angles. The removal of the  disrupted electron beams is a 
challenging 
problem and can be solved in simplest way in the crab crossing scheme of 
beam collisions.

The effects limiting the luminosity of \GG\ collisions are
different from those in  \EPEM\ collisions; the
beamstrahlung and instabilities are absent and only the coherent
\EPEM\ pair production in the field of oppositely moving ``used''(spent)
electron beam is still important. Therefore the electron beams for \GG\ 
collisions may have the smaller horizontal size at the IP  than in
\EPEM\ case, and consequently the larger luminosity 
is attainable.
 To reach ultimate luminosity in \GG\ collisions, the beams with
smaller horizontal emittance are required, which demands reoptimization
of the damping rings or use of low emittance RF guns. The latter
opportunity seems promising.

The laser required for conversion must be in \MKM\ wave length region,
with few Joules flash energy, about one picosecond duration and $20-30$~kW
average power. Such a laser can be a solid state laser with diode
pumping and chirped pulse amplification. These parameters can be
obtained also with a one pass free-electron laser. For 1--2 \TEV\ c.m.s.
energy collider, the optimum wave length is 2--4 \MKM\ and here FEL is
the only choice. The optical mirror system in tight space inside the
detector which transports the laser beam into the conversion points is also
a challenging task.

The idea of producing \GE\ and \GG\ collisions at linear colliders via
Compton backscattering has been proposed and studied by scientists
from Novosibirsk \cite{GKST81,GKST83,GKST84,TEL90,TEL95}.  Photon
colliders were considered also in 
\cite{BAL82,SKR,KONDR82,%
SPENCER85,SENS,CHTEL,TEL91,TEL92,BBC,BORDEN93,TEL93,RICHARD,MILLER,%
BAL95,SAL95,SPENCER95,KIMLC,KIM95,TELSH,MILLERD,MILLERM,TELMOR,SHULT}.

 A review of \GG, \GE\ colliders, physics
opportunities and available technologies can be found in the
proceedings of a workshop on gamma-gamma colliders held at Berkeley
\cite{BERK} and in the Zero Design Report of NLC \cite{NLC}.

The physics program at photon colliders was discussed at Workshops on
Physics at Linear Colliders (Saariselka, 1991; Waikoloa, 1993,
Morioka-Appi, 1995) at Workshops on \GG\ Collisions (San Diego, 1992;
Sheffield, 1995), in reviews 
\cite{GINSD,BRODHAW,BAIL94,BRODBER,CHANBER,GINBER,GINSH}
and in numerous papers.

\section{Physics at \GG, \GE\ collider}

\subsection{Introduction}

In what follows we assume the following parameters of \GG\
and \GE\ collisions (we use abbreviation PLC -- Photon Linear
Collider -- for these modes). Here $W_{\GG} $ and
$W_{\GE}$ denote c.m.s. energies of \GG\ or \GE\ systems, respectively.

\begin{itemize}
\item One can vary both initial electron energy and laser
frequency (a Free Electron Laser seems preferable for
this goal) to have the most sharp possible spectrum of photons
with a peak at the necessary energy.
High energy photons will have high degree of longitudinal polarization.

\item The width of the high energy luminosity peak will be 
$\Delta W_{\GG} / W_{\GG} \approx 0.15$, \newline $\Delta
 W_{\GE}/W_{\GE} \approx 0.05$\ \footnote{Outside 
the high energy peak, usually there is a flat part of the luminosity
distribution with 2--10 times larger total luminosity depending
on the details of the collision scheme (see sect.4).}.   

\item The annual ($10^7$~s) \GG\ luminosity will be about 10--30 fb$^{-1}$
(in the high energy peak) with possible upgrade of luminosity by one
order of magnitude (sect.4.5.3).

\item The annual \GE\ luminosity will be about 15--50 fb$^{-1}$
with about 5\% mono\-chro\-ma\-ti\-city.
\end{itemize}
\subsection{ Higgs boson physics}

Discovery and study of Higgs boson(s) properties will be of primary
importance at future $pp$ and linear $e^+e^-$ and \GG\ colliders. The
survey of the Higgs physics opportunities of PLC is simultaneously a
very good example showing how the complete phenomenological portrait is
obtained only by combining the complementary information available from
these distinct types of machines.

\subsubsection{Measurements of the Higgs boson couplings.}

Using the \GG\ collider mode a unique possibility appears
\cite{BARKLOW,BBC,GunionHaber,BalG,BORDEN93,Wat} to produce the Higgs
boson as an $s$-channel resonance decaying, for instance, into $b\bar
b$:
$$
\gamma\gamma\to h^0 \to b \bar b.
$$ 
Assuming that 300-500~GeV linear collider will first start operating
in $e^+e^-$ mode, the mass of the $h^0$ will already be known from the
Bjorken reaction $e^+e^-\to Z^* \to Zh$, and we can tune the energy of
the \GG\ collider so that the photon-photon luminosity spectrum peaks
at $m_h$. The cross section at PLC is proportional to the product
$\Gamma(h\to\gamma\gamma)\cdot BR(h\to b\bar b)$. The Higgs two-photon
decay width is of special interest since it appears at the one-loop
level.  Thus, any heavy charged particles which obtain their masses
from electroweak symmetry breaking can contribute in the loop. The
branching ratio $BR(h\to b\bar b)$ will also already be known from
$e^+e^-$ annihilation. Indeed, measuring $\sigma(ZH)$ (in the missing
mass mode) and $\sigma(ZH)BR(h\to b\bar b)$ in $e^+e^-$ mode of the
linear collider we can compute
$$
BR(h\to b\bar b) = \frac{[\sigma(ZH)BR(h\to b\bar b)]}{\sigma(ZH)},
$$
the error in the branching ratio is estimated at $\pm(8\div 10)\%$
\cite{GunionMartin}.

Then measuring the rate for the Higgs boson production in \GG\ mode of
the linear collider (accuracy $\pm 5\%$) we can determine the value of
the Higgs two-photon width itself (accuracy $\pm(11\div 13)\%$)
\cite{GunionMartin}
$$
\Gamma(h\to \gamma\gamma) = 
\frac{[\Gamma(h\to\gamma\gamma)BR(h\to b\bar b)]}{BR(h\to b\bar b)}.
$$

The main background to the $h^0$ production is the continuum
production of $b\bar b$ and $c\bar c$ pairs. In this respect, the
availability of high degree of photon beams circular polarization is
crucial, since for the equal photon helicities $(\pm\pm)$ that produce
spin-zero resonant states, the $q\bar q$ QED Born cross section is
suppressed by the factor $m_q^2/s$ \cite{BBC}. Another potentially
dangerous backgrounds originate from the resolved-photon processes
\cite{EGGHZ,BBB,JikT} in which a gluon from the photon structure
function produces $b\bar b$, $c\bar c$ pairs, and from the continuum
production of $b\bar b$ pairs accompanied by the radiation of
additional gluon \cite{Khhiggs}, calculated taking into account large
QCD ${\cal O}(\alpha_s)$ radiative corrections \cite{JikT}, which are
not suppressed even for the equal photon helicities. However, these
detailed studies have shown that the Higgs signal can still be
observed well above the background with the statistical error of the
Higgs cross section at the 6-10\% level with 20~fb$^{-1}$ in the wide
range of Higgs mass $60\div 150$~GeV.  As the width $\Gamma(h\to
\gamma\gamma)$ computed in the presence of an extra generation with
$m_L=300$~GeV and $m_U=m_D=500$~GeV drops to 15-30\% of its SM value
\cite{GunionHaber} this accuracy is sufficient to exclude the
contribution of a heavy fourth generation at the $5\sigma$ level.

If the Higgs boson is in the intermediate mass range, as it is implied
by the MSSM, it is also most likely to be observed at the LHC in the
gluon fusion reaction in its \GG\ decay mode $gg\to h^0\to
\gamma\gamma$, so that the measured rate is proportional to $BR(h\to
gg)\cdot \Gamma(h\to\gamma\gamma)$ (with an error of order $\pm 22\%$
at $m_{h_{SM}}=120$~GeV \cite{LHC}). The observable cross section for
the \GG\ signal at the LHC can depend quite strongly on the masses and
couplings of the superpartners and Higgs bosons, particularly if they
are not too heavy, and it varies from a few fb to more than 100~fb
over the parameter space of the MSSM, even in the scenario that
supersymmetry is not discovered at LEP2 \cite{KKMW}. In general,
interpretation of this one number is ambiguous, however by combining
this number with the value of the Higgs two-photon decay width,
measured in \GG\ and $e^+e^-$ experiments one can calculate the
two-gluon branching ratio $BR(h\to gg)$. Moreover, by measuring in
$e^+e^-\to Zh$, $e^+e^-\to \nu_e\bar\nu_e h$ ($W^+W^-$-fusion)
reactions the event rates for $h\to\gamma\gamma$ and $h\to b\bar b$,
one can compute the two-photon branching ratio
\begin{eqnarray}
BR(h\to\gamma\gamma)& = &BR(h\to b\bar b)
\frac{[\sigma(Zh)BR(h\to\gamma\gamma)]}{[\sigma(Zh)BR(h\to b\bar b)]}
\nonumber \\
& = & BR(h\to b\bar b)
\frac{[\sigma(\nu_e\bar\nu_e h)BR(h\to\gamma\gamma)]}
{[\sigma(\nu_e\bar\nu_e h)BR(h\to b\bar b)]}
\nonumber
\end{eqnarray}
with the accuracy $\pm (20\div 30)\%$ \cite{GunionMartin} and,
finally, compute in a model-in\-de\-pen\-dent way the total and partial
Higgs decay widths that are directly related to fundamental couplings
$$
\Gamma^{tot}_h=\frac{\Gamma(h\to\gamma\gamma)}{BR(h\to\gamma\gamma)},\;\;\;
\Gamma(h\to b\bar b) =\Gamma^{tot}_h BR(h\to b\bar b), 
$$
$$
\Gamma(h\to gg) =\Gamma^{tot}_h BR(h\to gg).
$$

For the Higgs bosons heavier than $2M_Z$ the Higgs signal in \GG\
collisions can be observed in $ZZ$ decay mode \cite{GunionHaber,BBC1}
if one of the $Z$'s is required to decay to $l^+l^-$ to suppress the
huge tree-level $\gamma\gamma\to W^+W^-$ continuum
background. However, even though there is no tree-level $ZZ$
continuum background, such a background due to the reaction
$\gamma\gamma\to ZZ$ does arise at the one-loop level in the
electroweak theory \cite{Jikzz,Bergerzz,DicusKao} which makes the
Higgs observation in the $ZZ$ mode impossible for $m_h>(350\div
400)$~GeV. It was found that for $185<m_h<300$~GeV the $ZZ$ mode will
provide a 8-11\% determination of the quantity
$\Gamma(h\to\gamma\gamma)\cdot BR(h\to ZZ)$.

In the 150-185~GeV window, the $W^+W^-$ and $b\bar b$ Higgs decay
modes are comparable. A recent study \cite{MTZ} has looked at the
$W^+W^-$-channel by also taking into account the interference between
the $\gamma\gamma\to W^+W^-$ continuum and the $s$-channel Higgs
exchange, which are all of the same electroweak order on
resonance. Unfortunately, for the measurement of the interference
pattern in this reaction the energy resolution less than 1~GeV is
required that is practically impossible to reach at photon colliders.
So, the accuracy of the two-photon Higgs width measurement might not
be better than 20-25\% in this region.

If $M_H\sim 2M_t$, the interference between QED process $\gamma
\gamma \to t\bar t$ and resonant one  $\gamma \gamma \to H\to
t\bar t$ can be used to obtain the value of Higgs coupling with
$t$--quark \cite{BGMel}.

Because of the dominance of the $W$ loop contribution in the three
family case, the $h\gamma\gamma$ vertex is also very sensitive to any
{\bf anomalous couplings of the Higgs or $\boldmath W$ bosons}
$hW^+W^-$, $W^+W^-\gamma$, $h\gamma\gamma$
\cite{GHiggs,GinSD,GounarisRenard}.  The sensitivity to anomalies in
these couplings can be comparable to that provided by LEP2 data.

{\bf The anomalous $\boldmath Z\gamma H$ interactions} can be studied
via $e\gamma\to eH$ process with longitudinally polarized electrons
\cite{GIv96}.

Finally, the production of Higgs bosons at a \GG\ collider offers
a special experimental opportunity to determine the $CP$ properties of
a particular Higgs boson \cite{Grzad92,GunionKelly,ZER94}. If $\vec E$
and $\vec B$ are electric and magnetic field strengths, a CP-even
Higgs boson $H^0$ couples to the combination $\vec E^2-\vec B^2$,
while a CP-odd Higgs boson $A^0$ couples to $\vec E\vec B$. The first
of these structures couples to linearly polarized photons with the
maximal strength if the polarizations are parallel, the letter if the
polarizations are perpendicular:
$$
\sigma \propto (1\pm l_{\gamma 1}l_{\gamma 2}\cos 2\phi),
$$
where $l_{\gamma i}$ are the degrees of linear polarization and $\phi$
is the angle between $\vec{l}_{\gamma 1}$ and $\vec{l}_{\gamma2}$. The
$\pm$ sign corresponds to $CP=\pm 1$ scalar particle.  The attainable
degree of linear polarization $l_{\gamma}$ at PLC depends on the value
of $z_m=(W_{\GG})_{max}/2E_0$ which can be changed in the case of free
electron laser. For $z_m=0.82$ the degree of linear polarization is
$l_\gamma\sim0.33$ only, but $l_\gamma\ge 0.8$ at $z_m\le0.5$ that is
sufficient to study the CP quantum numbers of Higgs with
$m_H\le 250$~GeV at $2E_0=500$~GeV $ee$ collider.

Moreover, if the Higgs boson is a mixture of CP-even and CP-odd
states, as can occur {\it e.g.} in two-doublet Higgs models with
CP-violating neutral sector \cite{CPhiggs}, the interference of these
two terms gives rise to a CP-violating asymmetry in the total rate
for Higgs boson production for $(++)$ and $(--)$ helicities of the
initial photons \cite{Grzad92}
$$
{\cal A} = 
\frac{\sigma_{++}(\gamma\gamma\to H)-\sigma_{--}(\gamma\gamma\to H)}
{\sigma_{++}(\gamma\gamma\to H)+\sigma_{--}(\gamma\gamma\to H)}.
$$
Experimentally the measurement of the asymmetry is achieved by
simultaneously flipping the helicities of both of the initiating laser
beams. One finds \cite{Grzad92} that the asymmetry is typically larger
than 10\% and is observable for a large range of two-doublet parameter
space if CP violation is present in the Higgs potential.

\subsubsection{The discovery of the Higgs boson.}

In principle it might be possible to detect the Higgs boson at PLC for
$m_H$ somewhat nearer to $\sqrt{s_{ee}}$ than the $0.7\sqrt{s_{ee}}$
that appears to be feasible via direct $e^+e^-$ collisions
\cite{Gunion96}, as the full \GG\ c.m.s. energy $W_{\gamma\gamma}$
converts into the Higgs resonance, and the parameters of the initial
electron and laser beams can be configured so that the photon spectrum
peaks slightly above $W_{\gamma\gamma}=0.8\sqrt{s_{ee}}$. However, for
$\sqrt{s_{ee}}=500$~GeV the detection of the SM Higgs boson in the
range of greatest interest, which cannot be accessed by direct
$e^+e^-$ collisions, {\it i.e.} $m_H\ge 350$~GeV, is problematic
because of the large one-loop $ZZ$ background
\cite{Jikzz,Bergerzz,DicusKao} discussed above.

The PLC potential to discover Higgs bosons is especially attractive in
the search for heavy Higgs states in the extended models such as MSSM
\cite{GunionHaber,Gunion96}. The most important limitation of a
$e^+e^-$ collider in detecting the MSSM Higgs bosons is the fact that
they are produced only in pairs, $H^0A^0$ or $H^+H^-$, and the
parameter range for which the production process $Z^*\to H^0A^0$ has
adequate event rate is limited by the machine energy to $m_{A^0}\sim
m_{H^0}\le \sqrt{s_{ee}}/2-20$~GeV ($m_{H^0}\sim m_{A^0}$ for large
$m_{A^0}$) \cite{Gunion96}. At $\sqrt{s_{ee}}=500$~GeV, this means
$m_{A^0}\le 230$~GeV. As $e^+e^-\to H^+H^-$ is also limited to
$m_{H^\pm}\sim m_{A^0}\le (220\div 230)$~GeV, it could happen that
only a rather SM-like $h^0$ is detected in $e^+e^-$ mode of the linear
collider, and none of the other Higgs bosons are observed. On the
other hand, $H^0$ and $A^0$ can be singly produced as $s$-channel
resonances in the \GG\ mode and PLC might allow the discovery of the
$H^0$ and/or $A^0$ up to higher masses
\cite{GunionHaber,Gunion96}. Particularly interesting decay channels
at moderate $\tan\beta$ and below $t\bar t$ threshold are $H^0\to
h^0h^0$ (leading to a final state containing four $b$ quarks) and
$A^0\to Zh^0$. These channels are virtually background free unless
$m_h^0\sim m_W$, in which case the large $\gamma\gamma\to W^+W^-$
continuum background would have to be eliminated by
$b$-tagging. Discovery of the $A^0$ or $H^0$ up to about
$0.8\sqrt{s_{ee}}$ would be possible. For large $\tan\beta$, the
detection of the $A^0$ or $H^0$ in the $b\bar b$ channel should be
possible for masses $\le 0.8\sqrt{s_{ee}}$
\cite{GunionHaber,Gunion96}, provided that effective luminosities as
high as 200~fb$^{-1}$ can be accumulated  (Sect.~4.5.3).

\subsection{Gauge boson physics}

Without the discovery of a Higgs boson at LEP2, LHC or linear
collider, the best alternative to study the symmetry breaking sector
lies in the study of the self-couplings of the $W$.  The PLC will be
the dominant source of the $W^+W^-$ pairs at future linear colliders
due to the reaction \ggww\ with the large cross section, that fast
reaches at high energies its asymptotic value $\sigmaw =8\pi\alpha^2
/\mww \approx 81$~pb \cite{GKPS}, which is at least an order of
magnitude larger than the cross section of $W^+W^-$ production in
$e^+e^-$ collisions, see Fig.~\ref{fig3}. With the rate of about 1--3 
million of $W$ pairs
per year PLC can be really considered as {\bf a $\boldmath W$ factory}
and an ideal place to conduct precision tests on the anomalous triple
\cite{27,BAIL94} and quartic \cite{BeBu,BAIL94,BB} couplings of the
$W$ bosons.  In conjunction with $e^+e^-\to W^+W^-$ one can reach much
better precision on these couplings. In addition, in the process of
triple $WWZ$ vector boson production it is possible to probe the
tri-linear $ZWW$ and quartic couplings
\cite{BB,eboli,BAIL94,gamma-gamma2} as well as the $C$ violating
anomalous $ZWW$, $\gamma ZWW$ interactions \cite{gamma-gamma2}.  With
the natural order of magnitude on anomalous couplings
\cite{Fawzi-Talks}, one needs to know the SM cross sections with a
precision better than 1\% to extract these small numbers.  The
predictions for $W$ pair production, including full electroweak
radiative corrections in the SM are known with very little theoretical
uncertainty at least for energies below 1~TeV \cite{DennerWW,JikiaWW}.

The process of $W$ production with the highest cross section 
in \GE\ collisions, \gewnu, with the asymptotic cross section 
$\sigma_{\gewnu} = \sigmaw /8 \sww \approx$ 43 pb \cite{GKPS}, is very
sensitive to the admixture of right--handed currents in W coupling
with fermions, as $\sigma_{\gewnu} \propto (1-2 \lel)$. This process was
studied in details in refs. \cite{DennerWnu}, where the radiative
corrections were taken into account.

\begin{figure}[!hbt]
\centering
\scalebox{1.}[0.75]{
\includegraphics[width=5.3in, angle=0, bb=50 50 520 630,clip] {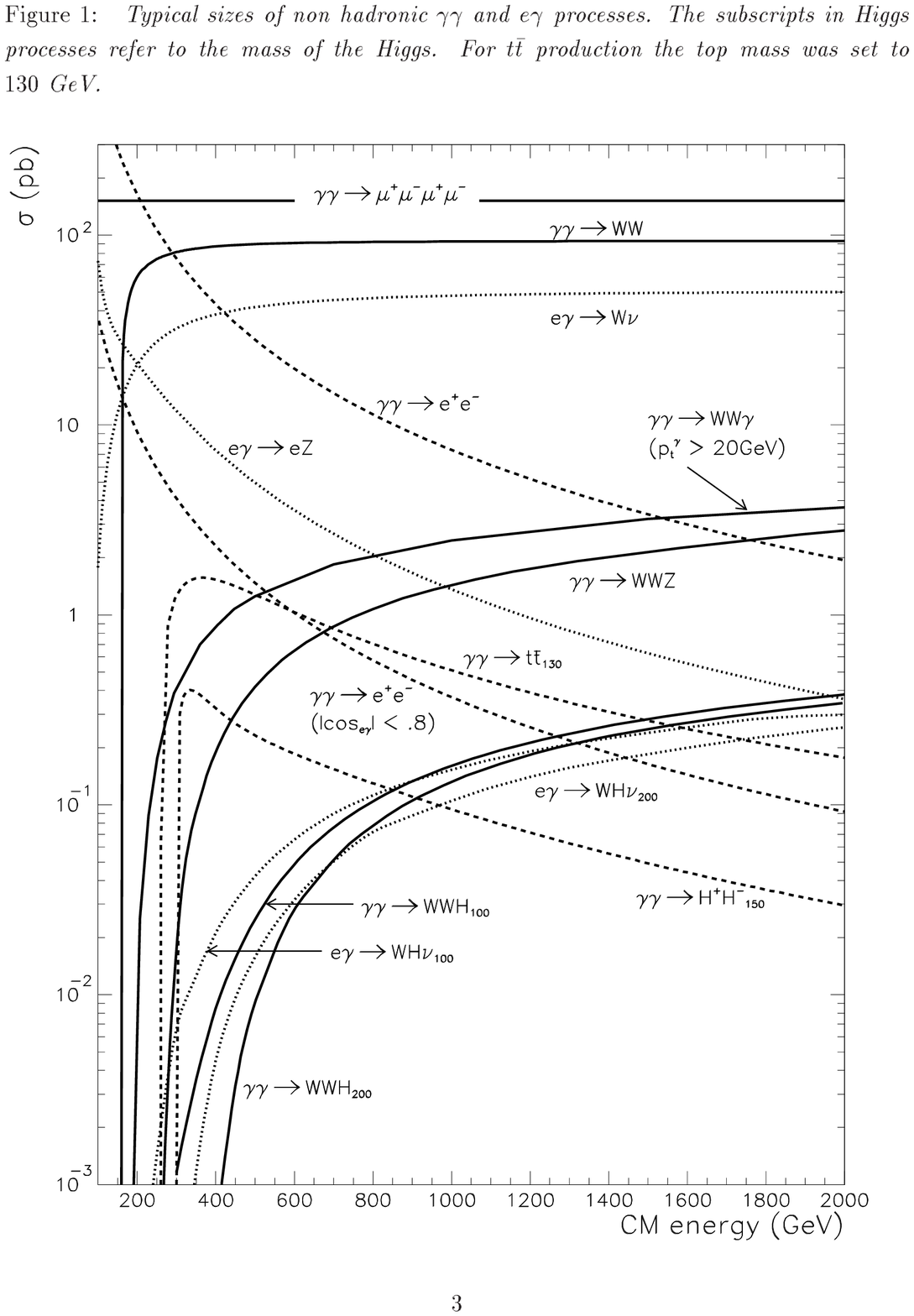}}
\caption{ The cross sections of some processes in \ggam and
$\gamma$e collisions.}
\label{fig2}
\end{figure}

\begin{figure}[!htb]
\centering
\includegraphics[width=3.in, angle=0, trim=50 130 50 50] {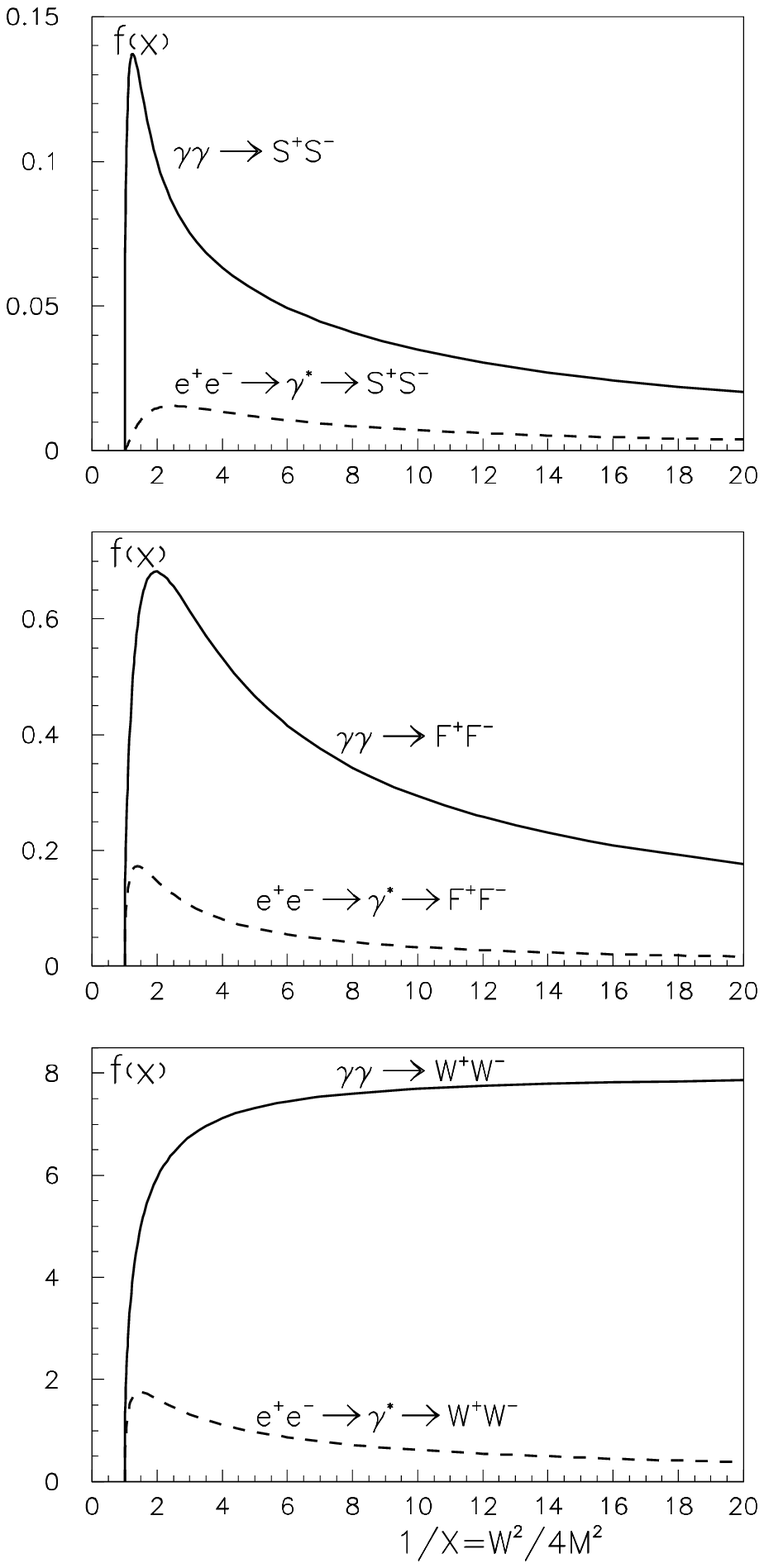}
\caption{ Comparison of  cross sections for charged pair production
in \EPEM\ and \GG\ collisions. The cross section  $\sigma
=(\pi\alpha^2/M^2)f(x)$, P=S (scalars), F (fermions),
W (W-bosons); M is particle mass, $x=W_{p\bar{p}}^2/4M^2$. The functions
$f(x)$ are shown.}
\label{fig3}
\end{figure}

The list of main processes with the W and Z production at PLC within
SM can be found in refs. \cite{Wrev,BBG}, see Fig.~\ref{fig2}.  When the
energy increases, the cross sections of a number of {\bf higher--order
processes} become large enough. The catalogue of such processes of
third order in SM is given in ref.  \cite{CompOur}.  It is will even
be possible to study at PLC pure one-loop induced SM reactions like
elastic light-by-light scattering $\gamma\gamma\to \gamma\gamma$
\cite{AAAA}, $\gamma Z$ \cite{AAAZ} or $ZZ$
\cite{Jikzz,Bergerzz,DicusKao} pair production processes, all of which
are dominated by the $W$ loop contribution at high energies.

At higher energy the effective $W$ luminosity becomes substantial
enough to allow for the study of {\bf $\boldmath W^+W^-\to W^+W^-$,
$\boldmath ZZ$ scattering} in the reactions $\ggam\to WWWW$, $WWZZ$,
when each incoming photon turns into a virtual $WW$ pair, followed by
the scattering of one $W$ from each such pair to form $WW$ or $ZZ$
\cite{BAIL94,BRODHAW,JikiaWWWW,CheungWWWW}. The result is that a
signal of SM Higgs boson with $m_H$ up to 700~GeV (1~TeV) could be
probed in these processes at 1.5~TeV (2~TeV) PLC, assuming integrated
luminosity of 200~fb$^{-1}$ (300~fb$^{-1}$).  The {\bf
$\boldmath\gamma Z\to WW$ scattering} can be studied in the \geeww\
reaction with the cross section of about 5~pb at $W_{\egam}\approx
0.4$~TeV and 27~pb at $W_{\egam}\approx 2$ TeV \cite{geeww,GIl96}.

\subsection{Physics of t--quarks}

The threshold effects will be investigated at PLC mode \cite{BGK,FKK}
similar to those at \epe\ mode \cite{FKhoz}. The strong dependence on
photon helicities provides opportunity to see delicate details of
\pair{t} interaction near the threshold, thought the broader momentum
spread of the photon beams do not allow to measure the exitation curve
with the same precision as in \EPEM\ energy scan.

The PLC provides the best opportunity for study of t--quark properties
themselves. Indeed, the cross section of $t\bar t$ production in the
\ggam collision is larger and it decreases more slowly with energy
than that in \epe collision (Fig.~\ref{fig3}).  Therefore, relatively
far from the threshold one can expect at PLC about $10^5$ $t\bar t$
pairs per year, their decay products being overlapped weakly. Some
rare $t$--decays could be studied here.

As mentioned in ref. \cite{kilgoremurayama}, the PLC has a great
advantage over $e^+e^-$ collider for the study of heavy scalar
superpartners such as the top squark, $\tilde t$. In $e^+e^-$
collisions $\tilde t \bar{\tilde t}$ pairs would be produced in the
kinematically suppressed $p$-wave and could not be effectively studied
unless the available collider energy were much greater than the
threshold production energy. In $\gamma \gamma$ collisions
stop-antistop pairs are produced in the $s$-wave, which can be further
enhanced by choosing photon beams of equal helicity.

\subsection{The new physics}

Two opportunities should be mentioned when we speak about effects of
new physics --- the discovery of new particles and new interactions of
known particles.  PLC provides the best place for discovering many new
particles in comparison with other colliders of the same energy. It is
connected with the following reasons:

{\it (i)} The signal to background (S/B) ratio at PLC is often
much better than that at hadron colliders (for more details  see
ref.  \cite{GinLBL}).

{\it (ii)} In comparison with hadron colliders  photons are
 "democratic" with respect to all charged particles. 

{\it (iii)} The cross sections of charged particles production in
\ggam collisions are larger than those in \epe collisions (see
Fig.~\ref{fig3}). Even if PLC luminosity is 5 times less than that for
basic \epe linear collider, the number of pairs produced at \epe
collider is not greater than at \ggam collider.  Besides, these cross
sections in \ggam collision decrease slower with growth of the
energy. This provides an opportunity to study new particles relatively
far from threshold with a substantial rate. In this region the decay
products of these new particles are overlapped weakly.  Therefore,
their detailed study will be more feasible.

{\it (iv)} $\gamma e$ collisions offer additional opportunities, such
as excited electron or neutrino production $\egam\to e^*$, $\egam
\to\textrm{W}\nu_e^*$, or single selectron production reaction $\egam
\to \tilde \gamma \tilde e$.

The physics potential of PLC in discovering of new particles and
interactions was considered in numerous papers (see also refs.
\cite{26}, \cite{DESY93} --\cite{SLAC96}). These include:

\begin{itemize}
\item SUSY particles \cite{Cuyp}:
$$
\GE \to \tilde{\W}\tilde{\nu};\;\;
\GG \to \tilde {\W^+}\tilde {\W^-};\; \GE \to \tilde{\Z}\tilde{e};\;\;
\ggam \to \tilde{P}\tilde{P},\;
(\tilde{P}\equiv\tilde{\ell},\;\tilde{H^{\pm}},\;\tilde{u},\;\tilde{d}).
$$
\item Excited leptons and quarks \cite{GIvBoud}:
$$
\egam\to e^*;\qquad
\ggam\to\ell^*\bar{\ell}; \qquad
\egam\to \nu^* \W;\qquad \ggam \to q^*\bar{q}.
$$
\item Leptoquarks \cite{lq}.
\item Charged Higgses.
\item Composite scalars and tensors.
\item Dirac--Schwinger monopoles with mass $\lesssim 10 E$ \cite{GPmon}.
\item Invisible axion (in the conversion region) \cite{Pol,GKP}.
\item Higgs nonstandard interactions \cite{GHiggs,GounarisRenard}.
\item The possibility to detect $\tilde{e}$\ in the process $\GE \to
\tilde{\Z}\tilde{e}$ with a mass higher than in \EPEM\ collisions
(where $\tilde{e}^+\tilde{e}^-$\ are produced in pairs), where
$\tilde{\Z}$\ is the lightest neutralino and $\tilde{e}$\ decays into
$\tilde{\Z}\tilde{e}$.
\end{itemize}

\subsection{ Hadron physics and QCD}

Hadron physics and QCD are the traditional fields for the \ggam\
collisions. The \ggam\ experiments provide new type of collisions with
the simplest quark structure of the pointlike initial state. The PLC
will continue these studies to new regions. The results from PLC
together with those from the Tevatron and HERA, will produce the
entire set of data related to a factorized (in the old Regge sense)
set of processes. In this respect, HERA gets a new importance of a
bridge between PLC and Tevatron/LHC (see review \cite{Eng} for some
details).

The expected values of {\bf the total cross section
$\boldmath\sigma(\gamma\gamma\to hadrons)$ and diffraction like
processes in the soft region} are:
$\sigma^{tot}\equiv\sigma_{\gamma\gamma\to hadrons} \sim 0.3\;\mu b$
in the SLC energy region, and $\sigma^{tot} \sim 0.5$--$1\;\mu b$ at
$W_{\ggam} \sim 2$ TeV \cite{CBP}.  Besides,
$\sigma(\ggam\to\rho^0\rho ^0)\sim 0.1\sigma^{tot}$ (see
\cite{BGMS}). The energy dependence of this cross section (together
with the $Q^2$ dependence in $\gamma e$ collisions) in comparison to
$\sigma_{pp} (\sigma_{p \bar p})$ and $\sigma_{\gamma p}$ will allow
us to understand the nature of the growth of hadron cross sections
with energy. The crucial problem is to test the possible factorization
of these cross sections (this factorization is assumed in
ref. \cite{CBP}).

{\bf For semihard processes} the nontrivial results in pQCD could be
obtained almost without any model assumptions due to the simple
pointlike nature of photons. The influence of hadronlike component of
photon is expected to be relatively small at large enough \ptr
only. For example, for the diffraction like processes it is expected
to be at $\ptr> 7$ GeV \cite{GI96}.

The processes $\ggam \to\rho^{0}X,\;\ggam\to\gamma X, \; \ggam\to
\rho^0\phi$ with rapidity gap are described by pure Pomeron
exchange. They present the best opportunity to study Pomeron.  The
processes $\ggam \to\pi^{0}X,\;\ggam\to\pi^0a_2$ with rapidity gap are
described only by odderon exchange. They present a unique opportunity
for odderon study.  The cross sections of some processes, integrated
over the range \footnote{ It corresponds to the production angle above
70-100 mrad at PLC with $W_{\gamma \gamma} = 100$--$180$ GeV.} of
$\ptr>7$~GeV and with large enough rapidity gap, are estimated from
below as \cite{GPS,GIv}.:
$$ 
\sigma_{\ggam\to \rho^0 X}\gtrsim 1\,{\rm 
pb},\quad\sigma_{\ggam\to \gamma X}\gtrsim 0.2\, 
pb,\quad\sigma_{\ggam\to\pi^0 X} \gtrsim 0.4\,{\rm pb}\,.
$$
The first two quantities should be multiplied by the
growing BFKL factor (see \cite{Muel,Iv}).

{\bf Photon structure function} is studied now at \epe\ colliders.  At
PLC it will be studied in a new region and with high accuracy (see
\cite{MilV}).

\section{Conversion region}
\subsection{Optimization of laser parameters, conversion efficiency}

The generation of high energy $\gamma$-quanta by Compton scattering of the
laser light on relativistic electrons is a well known method \cite{ARUT} and
has been used in many laboratories. However, usually the conversion
efficiency of electron to photons $k=N_{\gamma}/N_e$ is very small,
only about $10^{-7}$--$10^{-5}$. At linear
colliders, due to small bunch sizes one can focus the laser more tightly to the
electron beams and get $k\sim 1$ at rather moderate laser flash energy.
The kinematics of Compton scattering for this case, cross
sections, calculation of the conversion efficiency and consideration
of various processes in the conversion region (pair creation, nonlinear
effects) can be found in \cite{GKST83,GKST84,TEL90,TEL95}.

In the conversion region a laser photon with the energy $\omega_0$
scatters at a small collision angle on a high energy electron with the
energy $ E_0$.  The energy of the scattered photon $\omega$ depends on
its angle $\vartheta$ relative to the motion of the incident electron
as follows:

\begin{equation}
\omega = \frac{\omega_m}{1+(\vartheta/\vartheta_0)^2},\;\;\;\;
\omega_m=\frac{x}{x+1}E_0; \;\;\;\;
\vartheta_0= \frac{mc^2}{E_0} \sqrt{x+1};
\end{equation}
where
\begin{equation}
x=\frac{4E \omega_0 }{m^2c^4}
 \simeq 15.3\left[\frac{E_0}{\TEV}\right]
\left[\frac{\omega_0}{eV}\right] = 
 19\left[\frac{E_0}{\TEV}\right]
\left[\frac{\mu m}{\lambda}\right],
\end{equation}
$\omega_m$ is the maximum energy of scattered photons (in the direction of the
electron, Compton `backscattering').

For example: $E_0$ =250\,\, GeV, $\omega_0 =1.17$ eV ($\lambda=1.06$
\MKM) (Nd:Glass laser) $\Rightarrow$ x=4.5 and $\omega/E_0 = 0.82$.

The energy of the backscattered photons grows with increasing of
$x$. However, at $x > 2(\sqrt{2}+1)\approx 4.8$~\cite{GKST83}, 
high energy photons are
lost due to \EPEM\ creation in the collisions with laser photons. The
maximum conversion coefficient (effective) at $x\sim10$ is $k_{max}\sim$
0.3 \cite{TEL90}, while at $x < 4.8$ it is about 0.65 (one
conversion length). The luminosity in the latter case is
increased  by a factor 5. For $x=20$ and $x=4.8$ the difference 
is already of one order of magnitude.  
Further we will consider only the case $x \sim 4.8$, though
higher $x$ are also of interest for the experiments in which the ultimate
monochromaticity of \GG\ collisions is required.

 The wave length of the laser photons corresponding to $x=4.8$ is
\begin{equation}
 \lambda= 4.2 E_0 [\TEV]\;\; \mu m.
\end{equation}
For $2E_0 = 500\;$ \GEV\ it is about 1 \MKM, that is exactly
the region of the most powerful solid state lasers.

\begin{figure}[!hbp]
\centering
\vspace*{3mm}
\includegraphics[width=3in,angle=0,trim=100 260 30 50]{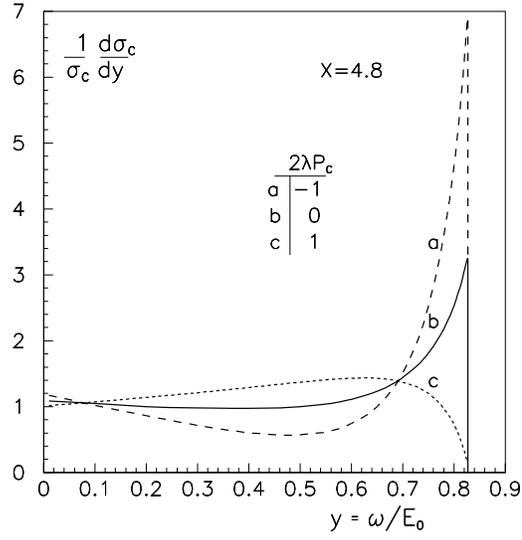}
\caption{Spectrum of the Compton scattered photons for different
polarizations of the laser and electron beams.}
\label{fig4}
\end{figure}

The energy spectrum of the scattered photons for $x=4.8$ is shown in
Fig.~\ref{fig4}  for various helicities of electron and laser beams
(here $\lambda_e$ is the mean electron helicity 
($|\lambda_e| \leq 1/2$), 
$P_c$ is the  helicity of laser photons. We see that with the 
polarized beams at $2\lambda_eP_c = -1$ the number of high energy 
photons nearly doubles. This case we will use in further examples.  
 The photon energy spectrum presented in Fig.~\ref{fig4} corresponds 
to the case of small conversion coefficient.  In the ``thick'' target each
electron may undergo multiple Compton scatterings~\cite{TEL95}. The secondary
photons are softer in general and populate the low  part of the spectrum.  

\begin{figure}[!hbp]
\centering
\vspace*{5mm}
\includegraphics[width=3in,angle=0,trim=100 260 30 50]{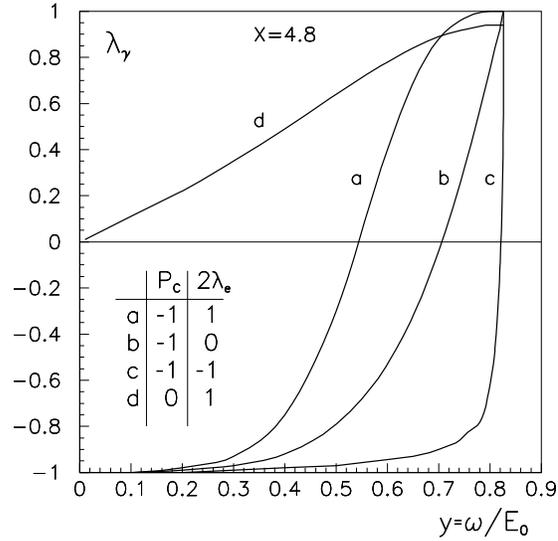}
\caption{Mean helicity of the scattered photons.}
\label{fig5}
\end{figure}

The mean helicity of the scattered photons is shown in Fig.~\ref{fig5}
for various helicities of the electron and laser beams~\cite{TEL95}.  For 
$ 2P_c\lambda_e = -1$ (the case with good monochromaticity) all  photons 
in the high energy peak have a high degree of like-sign polarization. If the
electron polarization is less than 100\% and $|P_c|= 1$, nevertheless
the helicity of the most high energy photon is still 100\% , but the energy
region with a high helicity becomes narrower. Low energy photons are
also polarized, but due to contribution of multiple Compton scattering
this region is not useful for the polarization experiments.  
The most valuable region for experiments is that 
near the maximum photon energy.
Higher degree of longitudinal photon polarization is essential 
for suppression of the QED background in the search (and study) of
the intermediate Higgs.
Note that at 0.5 TeV $ee$ collider the region of the intermediate 
Higgs  can be studied with rather small $x$. 
In this case the helicity of scattered photons is 
almost independent on polarization of the electrons, and 
if $P_c=1$ the high energy photons have very high circular polarization
in wide range near the maximum energy even with $\lambda_e=0$~\cite{TEL}. 

The measurement of CP-parity of the Higgs boson in \GG\ collisions can be 
done using linearly polarized photons (sect.2). At $y=y_m$ the degree
of linear polarization~\cite{GKST84}

\begin{equation}
l_\gamma=\frac{2}{1+x+(1+x)^{-1}},
\end{equation}

\noindent it is 33\% at $x=4.8$ and more than 80\% at $x\le1$.
Recently~\cite{KS96} it was suggested to transform circular polarization
of high energy photons into linear without loss of intensity using 
an additional linearly polarized laser bunch. For this method additional
studies are required.  

For the calculation of the conversion efficiency  it is
useful to remember the correspondence between the parameters of the
electron beam and laser beam: the emittance of the Gaussian laser
bunch with diffraction divergence is $\epsilon_{x,y} = \lambda/4\pi$,
the beta function at a laser focus $\beta\equiv Z_R$, where $Z_R$ is
known as the Rayleigh length in the optics literature. The r.m.s. spot
size of a laser beam at the focus ($i=x,y$)~\cite{GKST83} 
\begin{equation}
 \sigma_{L,i} = \sqrt{\frac{\lambda}{4\pi}Z_R}.
\end{equation}
The r.m.s. transverse size of a laser near the conversion region
depends on the distance $z$ to the focus (along the beam) as 
$$
\sigma_{L,x}(z)= \sigma_{L,x}(0) \sqrt{1+z^2/Z_R^2}.
$$
  We see that  the effective length of the conversion region is about $2Z_R$.

Neglecting multiple scattering, the dependence of the conversion
coefficient on the laser flash energy  A can be written as

$$
     k  = N_\gamma /N_e \sim 1-\exp (-A/A_0 ),
$$

\noindent
where $A_0$ is the laser flash energy for which the thickness of the 
laser target is equal to one Compton collision length.
This corresponds to $n_{\gamma}\sigma_{c}l = 1$, where
$ n_\gamma \sim A_0/(\pi \omega_0 a_{\gamma}^{2} l_\gamma )$,
$\sigma_c$ - is the Compton cross section ($\sigma_c =
1.8\cdot10^{-25}\;$ cm$^2$ at $x=4.8$), $l$ is the length of the region
with a high photon density,
which is equal to $2Z_R$ at $Z_R \ll \sigma_{L,z}\sim\sigma_z$
($\sigma_z$ is the electron bunch length). This gives
\begin{equation}
  A_0 = \frac{\pi\hbar c\sigma_z}{\sigma_c} = 5 \sigma_z [\MM],\,\,J
  \;\;\; \mbox{for}\;\; x=4.8.
\end{equation}

Note that the required flash energy decreases with reducing the
Rayleigh length up to $\sigma_z$, and it
is hardly  not changed with further decreasing of $Z_R$. This is
because the density of photons grows but the length having a high
density decreases and Compton scattering probability is almost constant.
 It is not helpful to make the radius of the laser beam at
the focus smaller than $\sigma_{L,x} \sim
\sqrt{\lambda\sigma_z/4\pi}$, which may be much larger than the
transverse electron bunch size at the conversion region.

For a more precise optimization of laser parameters we have used the
following formula for the conversion probability \cite{NLC}
$$ k = \frac{N_{\gamma}}{N_e} = 1-\frac{1}{\sqrt{2\pi}\sigma_z}
\int exp\;(-\frac{z^2}{2\sigma_z^2} - U(z))\;dz$$
\begin{equation}
  \mbox{where}\;\;U(z)=\frac{2\sigma_c A}{\sqrt{2\pi}\pi c \hbar Z_R
    \sigma_{L,z}} \int
  \frac{exp\;(-\frac{2(s-z/2)^2}{\sigma_z^2})}{1+s^2/Z^2_R}\;\;ds
\end{equation}
Here $\sigma_z, \sigma_{L,z}$ are the r.m.s. length of the electron and
laser beams respectively, $N_{\gamma}$ is the number of photons produced
by the electrons in their first Compton scatterings (which
can give photons with $\omega \sim \omega_{max}$).

For analyzing the conversion efficiency we have considered only the
geometrical properties of the laser beam and the pure Compton effect.
However, in the strong electromagnetic field at the laser focus,
multiphoton effects (non-linear QED) are important. 
Nonlinear effects are described by the parameter \cite{LANDAU,GKP}
\begin{equation}
\xi = \frac{eF\hbar}{m\omega_0 c},
\end{equation}
where $F$ is the r.m.s. strength of the electrical (magnetic) field in
the laser wave. At $\xi^2 \ll 1$ an electron interacts with one photon
(Compton scattering), while at $\xi^2 \gg 1$ an electron scatters on
many laser photons simultaneously (synchrotron radiation in a wiggler). 

\begin{figure}[!hbtp]
\centering
\vspace*{5mm}
\includegraphics[width=5in,angle=0,trim=30 20 30 30]{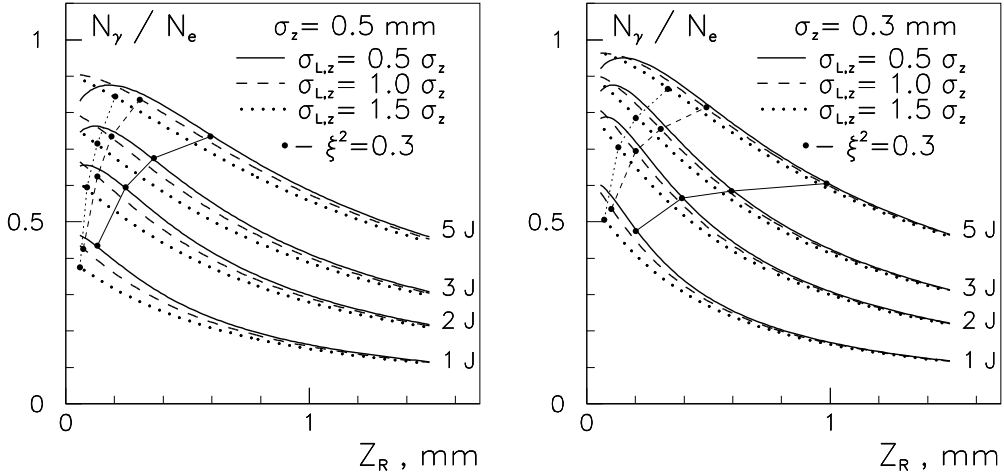}
\vspace*{5mm}
\caption{Conversion efficiency vs $Z_R$ for various values
of the flash energy and photon bunch length. Left figure
for electron bunch length $\sigma_z=0.5$~mm (TESLA), right 
for  $\sigma_z=0.3$~mm (SBLC).}
\label{fig6}
\end{figure}

The transverse motion of an electron in the electromagnetic wave leads
to an effective increase of the electron mass: $m^2
\rightarrow m^2(1+\xi^2)$, and the maximum energy of the scattered
photons decreases: $\omega_m = x/(1+x+\xi^2)$. At $x=4.8$ the value of
$\omega_m/E_0$ decreases by 5\% at $\xi^2=0.3$ \cite{TEL95}. This value
of $\xi^2$ we take as the limit. In the conversion region at $z=0$
\begin{equation}
\xi^2 = \frac{4r_e\lambda A}{(2\pi)^{3/2}\sigma_{L,z}mc^2 Z_R}
\end{equation}
The results of the calculation of $N_{\gamma}/N_e$ for various values of the
flash energy and beam parameters are presented in Fig.~\ref{fig6}. 
The points on the curves correspond to  $\xi^2 = 0.3$ (at $\lambda=1\;
\MKM$, optimum for $2E_0 = 500\;\GEV)$. These points limit the minimum
values of $Z_R$. Fig.~\ref{fig6}  may be used for other beam energies.
According to
eq.(6), $A_0$ depends only on the electron bunch length and  Compton
cross section (which is constant, if $x$ is kept constant). The value
of $Z_{R,min} \propto \lambda$ (see eq.(9))

\noindent
The requirements for the  laser parameters for $N_{\gamma}/N_e= 1 - e^{-1} =
0.632$ (one collision length ) are summarized in Table~\ref{table1}. The second
values of $Z_R$ correspond approximately to a 10\% drop in conversion
efficiency.

For the removal of disrupted beams at photon colliders it is 
necessary to use a crab crossing beam collision scheme
(see Fig.~\ref{fig9} in the next section). In this scheme the
electron beam is tilted relative to its direction of motion 
by an angle $\alpha=\alpha_c/2\sim15$~mrad. This means that
in the laser focus region the electron beam has an effective
size $\sigma_x=\sigma_z\alpha_c /2$, that is 7.5 and 4.5 $\mu$m
for TESLA and SBLC respectively. These sizes are comparable to,
and for TESLA even larger than  the laser spot size (see Table 1).
Therefore, for the equal conversion efficiency one has to
increase the laser flash energy by a factor of 1.9--1.5 respectively.
The other possible solution is the crab crossing collisions
of the electrons and laser bunches~\cite{XIE}.
The tilt of the  laser bunch can be obtained using ``chirped''
laser pulses and grating  as discussed in sect.7.1.
This solution is straightforward for solid state lasers, which
in any case use chirped pulses. For some schemes of free electron
laser this also can be done easily. For for simplicity we will take 
conservative values of the required flash energies: SBLC - 2 J, TESLA - 4 J.  

\begin{table}[!hbtp]

\caption{Required energy ($A_0$) and duration of laser flash
($\sigma_{L,z}$ for various length of the electron bunch ($\sigma_{z},$)
at $\lambda = 1\; \MKM, \;x=4.8$.
$ Z_R$ and $\sigma_{L,x}$ are optimum Rayleigh length and r.m.s. size of
the focal region.} 
\vspace*{0.4cm}
\begin{center}
\begin{tabular}{c|ccll} \hline
  $\sigma_{z},$ mm & $A_0,\;$J & $\sigma_{L,z}$, mm & $ Z_R$, mm & 
  $\sigma_{L,x}$, \MKM\   \\ \hline 
  0.3 & 1.5 & 0.3 & 0.15--0.2  & 3.5--4 \\ 
  0.5 &2.1  & 0.5 & 0.15--0.25 & 3.5--4.5 \\ 
  0.7 & 2.8 & 0.7 & 0.15--0.3  & 3.5--5 \\ 
\end{tabular}
\end{center}
\label{table1}
\end{table}

As follows from Fig.~\ref{fig6}  the nonlinear effects at $2E = 500\;\GEV\ 
(\lambda_{opt}=1\; \MKM)$ and $\sigma_z =$0.3--0.7 \MM\ have a small
influence to the required laser flash energy. For shorter electron
bunches and larger $\lambda$ the required flash
energy due to nonlinear effects may be much larger than deduced from the
diffraction consideration only \cite{TEL95}. Recently \cite{TEL96} it
was shown how the problem of nonlinear effects at the conversion
region can be avoided. Owing to non-monochromaticity of a laser light it
is possible to `stretch' the depth of a laser focus keeping the radius
of the focal spot size constant. In this scheme the required flash
energy is determined only by the diffraction and is given
approximately by  eq.(6).

\subsection{Low energy electrons after conversion.}

For the removal of the spent electrons it is important to know the values
of the maximum disruption angle and minimum energy of the spent electrons.
The disruption angles are created during beam collisions 
at the IP. The electrons with the lower energies have larger
disruption angles. The simulation code (to be described in the next
section) deals with about 5000 (initial) macroparticles and can not
describe the distribution tails. But providing the minimum energy and
energy dependence of the disruption angle are known, we can
correct the value  of maximum disruption angle obtained by the
simulation.

\begin{figure}[htbp]
\centering
\includegraphics[width=3in,angle=0,trim=30 30 30 80]{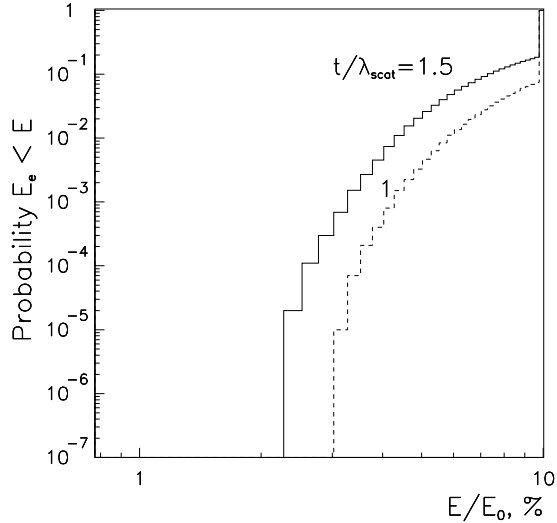}
\caption{Probability for
an electron after the conversion region to have an energy
below $E/E_0$.}
\label{fig7}
\end{figure}

Low energy electrons are produced at the conversion region due to
multiple Compton scatterings \cite{TEL90}. Fig.~\ref{fig7} 
shows the probability
that an electron which has passed the conversion region has the energy
below $E/E_0$. Two curves were obtained by simulation of $10^5$
electrons passing the conversion region with the thickness 1 and 1.5
of the Compton collision length (at $x=4.8$). Extrapolating these
curves (by tangent line) to the probability $10^{-7}$ we can obtain the
minimum electron energy corresponding to this probability: 2.5\%
and 1.7\% of $E_0$ for $t/\lambda_{scat}=$ 1 and 1.5 respectively. 
The ratio of
the total energy of all these electrons to the beam energy is about
$2\cdot10^{-9}.$ This is a sufficiently low fraction comparable
with other backgrounds (see sec.5).  So, we can conclude that 
the minimum energy of electrons after the conversion region is about 2\% of 
initial energy, in agreement with the analytical estimation \cite{TEL90}.

\section{Interaction region}

\subsection{Collision schemes}

\begin{figure}[!hbtp]
\centering
\includegraphics[width=3in,angle=-90]{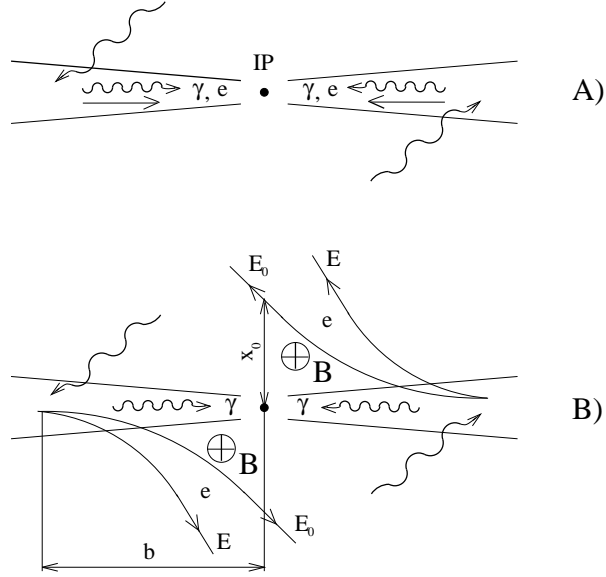}
\caption{Two basic collision schemes: a) ``without deflection'',
b) ``with deflection'' using the sweeping magnet.}
\label{fig8}
\end{figure}

We will consider two basic collision schemes (Fig.~\ref{fig8}):    

\underline{Scheme A (``without deflection'').} There is no magnetic
deflection of spent electrons and all particles after conversion
region travel to the IP \cite{TEL91,BAL95}. The conversion point may
be situated very close to the IP at the distance $b \sim$ 
5$\sigma_z$.

\underline{Scheme B (``with deflection'').} After conversion region
particles pass through a region with a transverse magnetic field where
electrons are swept aside \cite{GKST83,TEL90,TEL95}. Thereby one can achieve
a more or less pure \GG\ or \GE\ collisions.

The scheme A is simpler but background conditions are much worse and
disruption angles are larger. Additional background is connected not
only due to mixture of different types of collisions but also due to
emission of a huge number of beamstrahlung photons during beam collision. 
 This leads to ``background'' \GG\ and \GE\ luminosities at
small invariant masses exceeding the ``useful'' luminosity in the high
energy peak by one order. This causes an additional backgrounds to the
detector (see sect.5).

\begin{figure}[!hbtp]
\centering
\includegraphics[width=1in,angle=-90]{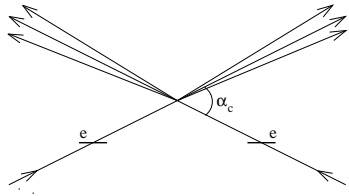}
\caption{Crab-crossing scheme.}
\label{fig9}
\end{figure}

In both schemes the removal of the disrupted spent beams can best be
done using the crab-crossing scheme \cite{PALMER} (Fig.~\ref{fig9}),
which is proposed in the NLC and JLC projects for \EPEM\
collisions. In this scheme the electron bunches are tilted (using RF
cavity) with respect to the direction of the beam motion, and the
luminosity is then the same as for head-on collisions. Due to the collision
angle the outgoing disrupted beams travel outside the final quads.
  In the next sections both schemes will be considered in detail.
\vspace*{0.5cm}

\subsection{Collision effects in \GG\ and \GE\ collisions.}
 
 Luminosity in \GG,\GE\ collisions is restricted by many factors:
\begin{itemize}
\item collision effects: coherent pair creation, beamstrahlung, beam
  dis\-pla\-ce\-ment);
\item beam collision induced backgrounds (large disruption angles of
  soft particles);
\item luminosity induced backgrounds (hadron production, \EPEM\ pair
  production);
\item problem of obtaining electron beams with small emittances.
\end{itemize}

For optimization of the photon collider it is useful to know
qualitatively the main dependences. In this section we will consider
collision effects which restrict the luminosity \cite{TEL90,TEL95}.

It seems, at first sight, that there are no collision effects in \GG\ 
and \GE\ collisions because at least one of the beams is neutral. This
is not completely correct because during beam collisions electrons and photons
are influenced by the field of the oppositely moving electron beam.
In \GG\ collisions it is spent beams, in \GE\ collisions the field is
created also by the ``main'' electron beam used for \GE\ collisions.

A strong field leads to the following effects.\\
 In \GG\ collisions:
conversion of photons into \EPEM\ pairs (coherent pair creation 
\cite{CHTEL}).\\
In \GE\ collisions: a) coherent pair creation; b) beamstrahlung;
c) beam displacement.

\vspace*{0.3cm}
{\it \underline{\GG\ collisions}}.\\
Coherent pair creation is exponentially suppressed for $\Upsilon =
\gamma B/B_0 \le 1\;\;(B_0 = m^2c^3/e\hbar = 4.4\cdot10^{13}$ Gauss).
But, if $\Upsilon > 1$, most of high energy photons can be converted to
\EPEM\ pairs during the beam collision~\cite{TEL90}. 
There are two ways to avoid
this effect (i.e. to keep $\Upsilon \le 1$):\\ 1) to use flat beams;\\ 2)
to deflect the electron beam after conversion at a sufficiently large
distance ($x_0$ for $E = E_0$) from the IP.

The condition $\Upsilon < 1$ corresponds approximately to the
condition \cite{TEL95} 
\begin{equation} x_0\;\mbox{or}\;\sigma_{x,min}
\ge 1.5\frac{Nr_e^2\gamma}{\alpha\sigma_z} \sim
30\left(\frac{N}{10^{10}}\right)\frac{E_0[TeV]}{\sigma_z[mm]},\;\NM\
\end{equation} For TESLA and SBLC $x_0,\sigma_{x,min} \ge
217(110)E_0[TeV]$ nm respectively.

In the scheme with flat beams, the minimum vertical beam size is
$\sigma_{\gamma,y} \sim b/\gamma$, where $b \ge 5\sigma_z$ 
(not 1-2$\sigma_z$\ because the field
in the conversion region should be small).  But, if $\sigma_y$ of the
electron beam is larger than this value, then it is reasonable to choose
$b/\gamma \sim \sigma_y$. This case is relevant for TESLA
and SBLC.
The maximum luminosity in this case is
\begin{equation}
  \LGG\ \sim \frac{k^2N^2f}{4\pi\sigma_{x,min}\sigma_{\gamma,y}} =
  \frac{k^2N\alpha\sigma_zf}{6\pi r_e^2\sigma_{\gamma,y}\gamma},
\end{equation}     
where $\sigma_{\gamma,y} = max(\sigma_y,5\sigma_z/\gamma)$.

If the available $\sigma_x$ is much smaller than $\sigma_{x,min}$
given by eq.(10), then it is reasonable to keep $\sigma_x$ as small 
 and to provide $\Upsilon < 1$ by deflecting beam on the distance 
$x_0$ given by eq.(10). The minimum photon spot size in this case is
$$
  \sigma_{\gamma,min} \sim b/\gamma \sim
  \sqrt{\frac{2E_0x_0}{eB_e}}\frac{1}{\gamma} \sim
  \sqrt{\frac{3Ner_e}{\alpha \sigma_zB_e}}  
$$
\begin{equation}
 \sim  7.5\left(\left[\frac{N}{10^{10}}\right]\left[
  \frac{mm}{\sigma_z}\right]\left[\frac{T}{B_e}\right]\right)^{1/2},\;\;\NM\,
\end{equation}
where $B_e$ is the transverse magnetic field in the region between the 
CP and IP. For ``nominal'' TESLA(SBLC) parameters (see sect 4.4)
 $ \sigma_{\gamma,min} \sim$ 20(14) nm at $B=1$~T.
Such photon spot can be obtained only when electron bunch sizes are
smaller than these values.
  The maximum luminosity in this case is
\begin{equation}
  \LGG\ \sim \frac{k^2N^2f}{4\pi \sigma_{\gamma,min} ^2} =
  \frac{k^2N\alpha\sigma_zf B_e} {12\pi e r_e}
\end{equation}  
However, with the horizontal emittances considered   
currently in the TESLA and SBLC projects the
level $\sigma_x \sim $ 15 nm is unreachable. In the case $\sigma_x >
\sigma_{\gamma,min}$
\begin{equation}
  \LGG\ \sim \frac{k^2N^2f}{4\pi \sigma_{\gamma,min} \sigma_x} =
  \frac{k^2N^2f} {4\pi \sigma_x}\sqrt{\frac{\alpha\sigma_z B_e}{3Ner_e}}
\end{equation}  

We have considered all cases important for optimization of \GG\ 
collisions.  For given attainable $\sigma_x$ and $\sigma_y$  one can
check which scheme gives larger liminosity. Final optimization should be done
by simulation. 
Note that in all cases, even for $\sigma_x > \sigma_{x,min}$, some
magnetic deflection is useful for reduction of backgrounds. 

All described above is valid for any collider parameters. But there is
one nice surprise: at $2E_0=500\;$\GEV\ and with the TESLA--SBLC
parameters there exists no coherent pair creation (due to beam
repulsion), and one can then obtain very high \GG\ luminosity which is
only determined by the attainable emittances of electron beam
\cite{TELSH} (for more detail see sect.4.5.3 ``Ultimate
luminosity'').

\vspace*{0.3cm}
{\it \underline{\GE\ collisions}}\\
In \GE\ collisions there are more collision effects: coherent pair
creation, beam\-strah\-lung and beam dis\-pla\-ce\-ment. The detail
consideration of these effects and calculation of the \GE\ luminosity
can be found elsewhere \cite{TEL95}. Briefly, the picture  is the
following.\\  
1) To avoid coherent pair creation the electron beams
should be flat, with $\sigma_x$ larger than it is given by eq.(10).\\
2) By choosing the distance $b > \gamma\sigma_y$ we can obtain very
monochromatic \GE\ collisions.\\ 
3) To maintain good monochromaticity, the
beamstrahlung losses of the main electron beam in the field of the
spent electron beam should be small. This can be provided by 
magnetic deflection of
the spent beam. \\ 
4) At low beam energies it can happen that due to
repulsion the ``main'' electron beam is shifted and does not collide with the 
high energy photons. Therefore, the magnetic
deflection of the spent electron beam should be kept large enough.

To meet all the above enumerated requirements the distance between the
CP and IP in \GE\ collisions must usually be larger than in \GG\
collisions, therefore the maximum attainable \GE\ luminosity is
somewhat smaller.
In the scheme without magnetic deflection one can also obtain a
sufficiently large \GE\ luminosity, but with much worse
quality of collisions.  The corresponding luminosity spectra are
obtained by  simulation in sect.4.5.

\vspace*{0.3cm}
{\it \underline{Disruption angles}}

The maximum disruption angle is an important issue for photon
colliders determining the value of the crab crossing angle.

One source of large angle particles is low energy electrons 
from the conversion region and the minimum energy of these electrons  is
about 0.02$E_0$ (sect.3.2).  In the scheme without magnetic deflection the soft
electrons are deflecting by the opposing beam by an angle \cite{TEL90}
\begin{equation}
\vartheta_d \sim 0.7\left(\frac{4\pi r_e N}
{\sigma_z \gamma_{min}} \right)^{1/2} \sim
 2\left(\frac{N/10^{10}} {\sigma_z [\MM\ ] E_0 [\TEV\ ]}
\right)^{1/2}\;\MRAD\ 
\end{equation} 
for $E_{min}=0.02E_0.$ 
The coefficient 0.7 here was found by tracking particles in the field of
the beam with a Gaussian longitudinal distribution for the TESLA--SBLC
range of parameters. For $2E_0$ = 500 \GEV\ for TESLA (SBLC)
$\vartheta_d = 9 (8)\; \MRAD\ $.

One can decrease $\vartheta_d$ by predeflecting the spent electron beam by the
external magnetic field. Unfortunately the kick angle (on the opposing beam)
changes very slowly up to
the displacement at IP $ \Delta y \sim
\sqrt{r_eN\sigma_z/\gamma_{min}}$. It is about 2.5 \MKM\ for the TESLA
and 1 \MKM\ for SBLC for $E_{min}=0.02E_0$.  For this impact distance
the kick angle decreases by a factor 1.7 only. At $b=1.5$ cm the
required $B_e$ is $ 3.7 (1.5)\;$~kGs for TESLA(SBLC). For larger fields 
the disruption angle decreases as $1/B_e$.

The second source of soft particles is hard beamstrahlung.  The
deflection angle is described by the same eq.(15), but with
coefficient 1.2 instead of 0.7, which corresponds to the case when
hard photon is emitted near the center of opposing beam. What is the
minimum energy of electrons after beamstrahlung? According to
Sokolov-Ternov formula the high energy tail is expressed as
$exp(-\xi_S)$, where $\xi_S =2y/(3\Upsilon(1-y))$ and $y=\omega/E$. As
soon as the total number of radiated photons is of the order of one,
this expression gives approximately the relative number of electrons
with the energy loss above $y$. The probability $10^{-7}$ corresponds
to $\xi_S = 16$.  For $\Upsilon = 1$, which is maximum for photon
colliders, we find $1-y_{max}=0.04$.  In other words the lowest energy
of electrons after beamstrahlung is about 4\% of $E_0$ (for $\Upsilon$
=1).

According to eq.(15) with the coefficient 1.2 the deflection angle of such 
electrons (for $\Upsilon=1$) will be larger by 20\% than that for low
energy electrons arising at the conversion region and 
$\vartheta_d \propto \sqrt{\Upsilon}$. 
The magnetic deflection decreases the disruption angles
of the low energy electrons coming from the conversion
region, however it does hardly  changes almost the disruption angles of soft
"beamstrahlung" electrons created by high energy electrons
which have small magnetic deflection and pass the IP close to the opposing
beam.
The presented picture of collision effects  helps to understand
numerical results of the  simulation.

\subsection{Simulation code.}

As we have seen, the picture of beam collisions in photon colliders is
so complicated that the best way to see a final result is a
simulation. In the present study we used the code written by
V.Telnov \cite{TEL95,LINCOL}.

 It is written for the simulation of  \EPEM, \EE, \GE, \GG\
beam collisions at linear colliders and takes into account the
following processes: 
\begin{enumerate} \item {\it Multiple Compton
scattering in the conversion region}.  In the
  present simulation ``good'' case of polarization $(2P_c\lambda_e
  =-1)$  was assumed and the thickness of the photon target was put
  equal to one conversion length. 
\item {\it Deflection by the external magnetic field and synchrotron radiation}
in the region between the CP and IP. 

\item {\it Electromagnetic forces, coherent pair creation and
beamstrahlung} during beam collisions at the IP.  

\item {\it Incoherent \EPEM\ creation in \GG, \GE, \EE\ collisions}.
\end{enumerate}

 Initial electron beams are described by about 3000 macroparticles
(m.p.) which have a shape of flat rectangular bars with sizes (x$\cdot$y)
equal to 0.4$\sigma_x\cdot0$. In the longitudinal direction the
electron bunch has a Gaussian shape ($\pm 3\sigma$) and is cut on 150 slices.
The macroparticles have only the transverse field and influence on
 macroparticles of the opposite bunch which have the same z-coordinate
(this coordinate changes  by steps). At initial positions macroparticles are
directed to the collision region in the way corresponding to beam
emittances and beta functions. During the simulation new macroparticles
(photons, electrons and positrons) are produced which are considered
further in the same way as the initial macroparticles.

  At the output the code gives all parameters of colliding pairs
(macroparticle "collides" when the distance between their centers is
less then $0.15\sigma_x$ on $x$ and less then $0.15\sigma_y$ on
$y$--directions) and all parameters of the final particles.
Incoherent \EPEM\ pairs are simulated separately after beam collisions.
         
The code was used for simulation of the NLC based photon
collider \cite{NLC} and the results are in agreement \cite{TAK} with
the code GAIN \cite{YOK} written later for the same purposes.

\subsection{Parameters of electron beams}

   The parameters of electron beams considered for \GG,\GE\ collisions
are presented in Table~\ref{table2}.

\begin{table}[thbp]
\caption{Parameters of electron beams.}
\vspace*{0.5cm}
\begin{center}
\begin{tabular}{ccccccc} \hline
 &{\scriptsize TESLA(1)} &\scriptsize TESLA(2) &\scriptsize TESLA(3) & 
  \scriptsize TESLA(4) &\scriptsize SBLC(1) & \scriptsize SBLC(2) 
    \\ \hline 
\normalsize
$N/10^{10}$& 3.63 & 1.82 & 3.63 & 3.63 & 1.1 & 1.1 \\  
$\sigma_{z}$, mm& 0.5 & 0.5 & 0.5 & 0.5 & 0.3 & 0.3 \\  
$f_{rep}$, Hz& 5 & 4 & 5 & 5 & 50 & 50 \\
$n_b$& 1130& 2260 & 1130 & 1130 & 333 & 333 \\
$\Delta t_b$, ns  & 708 & 354 & 708 & 708 & 6 & 6 \\ 
$\gamma \epsilon_{x,y}/10^{-6}$,m$\cdot$rad & $14.,0.25$ & $12.,0.03$ & 
$4.,0.25$&  $1.,1.$ & $5.,0.25$ & $0.5,0.5$ \\
$\beta_{x,y}$,mm at IP& $3.2,0.5$ & $3.2,0.5$ & $2.0,0.5$ &
$1.1,0.5$& $2.5,0.4$ & $0.77,0.3$ \\
$\sigma_{x,y}$,nm& $303,16$ & $280,5.5$ & $128,16$ & 
$47,32$ & $160,14$ & $28,17.5$ \\  
$L(geom),10^{33}$& 12.2 & 15.4 & 30.0 & 39 & 7.1 & 32.6\\ \hline  
\end{tabular}
\end{center}
\label{table2}
\end{table}

 Some comments: \\ TESLA(1) --- is the basic variant considered for
\EPEM\ collisions at the TESLA. \\ TESLA(2) --- is the second basic
TESLA set of parameters for \EPEM\ collisions with reduced vertical
emittance. \\ TESLA(3) --- as TESLA(1), but the horizontal emittance is
reduced by a factor of 3.5; that may be achieved with optimized damping
rings. \\ TESLA(4) --- the variant with a low emittance polarized
RF-gun (without damping rings). The emittances used here are by a
factor of 5 lower than presently achieved for this number of
particles~\cite{TRAV}. However, it seems possible to join 
(using some difference
in energies) many (5-10) low current beams with low emittances to one
beam with the number of particles as in TESLA (1,3) variants. \\
SBLC(1) --- is the basic variant considered for \EPEM\ collisions at
the SBLC. \\ SBLC(2) ---- the variant with a low emittance polarized
RF-guns (see comment for TESLA(4))

   For demonstration of ultimate parameters of the photon collider
based on the TESLA we will consider also the ``super'' variant TESLA(4),
where in comparison with TESLA(4) the emittances are further reduced by a
factor 5. Such emittance of electron beams can be achieved using the 
method of laser cooling proposed recently \cite{TEL96}. This method
requires lasers with the flash energy  about 10 J (by a factor 3-4 larger than
for e$\rightarrow\gamma$ conversion) which seems possible in the scheme having
a laser photon recirculation. In this proposal we do not consider this option
in the ``main list''. 
  
  As was noted before the electron beams in \GG\ collisions can have
smaller horizontal beam sizes than in \EPEM\ collisions. Beta
functions presented in Table 2 are minimum for given beam parameters.
In some case they are larger than $\sigma_z$ due to Oide effect
(chromatic aberrations due to synchrotron radiation in the final
quads). We assumed that the final focusing system has the same
structure as for \EPEM\ collisions and the distance between IP and the
nearest quad is 2 m. 

  Due to beam collisions at large collision angle the final quads
should have a special design with a ``hole'' for disrupted
beams. Moreover, the magnetic field in this region should be kept
small enough ($<0.05$ T), so that particles with the lowest energy
(about 5 GeV) get a small deflection and follow essentially in the
direction of the beam dump.

One  possible solutions is to use a
superconducting ironless quadrupole with concentric current of opposite
polarities proposed for NLC \cite{NLC}. The outer radius of the coils
in this design is 3 cm.  So, the minimum crossing angle is about 4/200
rad + disruption angle ($\sim$ 10 mrad), that is about 30 mrad.
This is the value of crab crossing angle we will use in our
considerations.
  Others quad designs with iron poles are also possible. Low field
region in this case can be arranged inside the quad using magnetic screens.
 
\subsection{Simulation results}
\subsubsection{Scheme without deflecting magnets}

   In the scheme without sweeping magnets  
there is only one free parameter: the distance between the 
CP and IP. It is reasonable to take $b=1.5\gamma\sigma_y$. For such
$b$ the spectral luminosity for hard photons is almost the same as for
$b=0$, but the low energy part of \GG\ luminosity is suppressed (because
the spot size for low energy photons at the IP is larger than that for high
energy photons).  With this choice the value of $b$ varies between 0.4--2.4 cm
for the considered beam parameters.  It was assumed that the thickness of laser
target is equal to one collision length (for electrons with the 
initial energy) that corresponds to  $k=0.63$. 

\begin{figure}[!hbtp]
\centering
\includegraphics[width=5.2in,angle=0,trim=30 20 30 30]{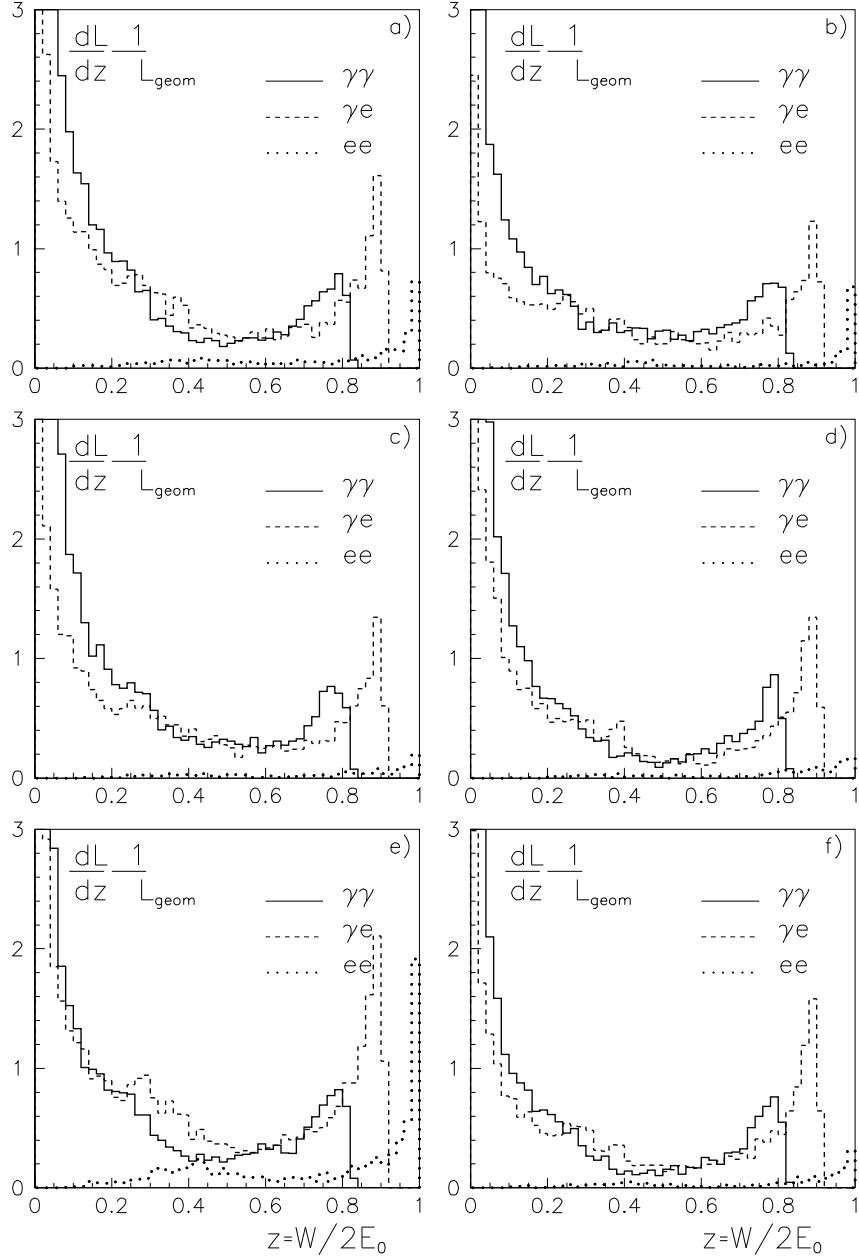}
\caption{Luminosity spectra for the scheme \underline{without deflection} 
for the \GG\ collider parameters presented in 
Tables~\ref{table2},~\ref{table3} 
(Figs.~\ref{fig10}a), b)... correspond to TESLA(1), TESLA(2)...).
Luminosity distributions are normalized to the geometrical luminosity.
For additional comments see the text.}
\label{fig10}
\end{figure}

  The results of simulation for \GG, \GE\ and \EE\ luminosities are
presented in Fig.~\ref{fig10} and Table~\ref{table3} (the first lane).  
All luminosity distributions are normalized to the geometrical luminosity
$L_{geom} = N^2f/(4\pi\sigma_x\sigma_y)$. Looking to these data one
can see: 
\begin{itemize}
\item \LGE (total) $\sim$\LGG(total) and \LGE($z>0.65) \sim$
(1--2)\LGG($z>0.65$); so \GG, \GE\ collisions can be studied simultaneously; 
\item \LGG(total) $\sim$ 10\LGG($z>0.65$); 
      low energy \GG\ luminosity will give additional backgrounds; 
\item \LEE\ $<$ 0.1 \LGG; (due to beam repulsion); 
\item \LGG($z>0.65$) = (1.2--3.3)$10^{33}$ \CMS;   
\item   $N(\GG\to \textrm{hadrons} )/ \textrm{collision} 
\sim $0.1--2 with $\overline{W_{\GG}}\sim 0.15\times2E_0$.
\item maximum disruption angles $\vartheta_x,\vartheta_y < 10\;\MRAD$. 
\end{itemize}

We see that in the scheme without magnetic deflection there is mixture
of \GG, \GE, \EE\ collisions and most of collisions have small invariant
masses. This low energy luminosity is produced by the soft photons after
multiple Compton scattering and beamstrahlung photons created during
beam collision.

\subsubsection{Scheme with magnetic deflection}

\underline{\it \GG\ \it collisions}

\vspace{2mm}

 Magnetic deflection allows having pure \GG\ collisions and suppress
\GE,~\EE\ backgrounds. It is important to note also that deflection
 of spent electrons can significantly suppress low
energy \GG\ luminosity. The swept electron beam radiates many beamstrahlung
photons in the field of the opposing spent electron beam but now 
these photons are not collided with particles from the opposing beam.

\begin{figure}[!hbtp]
\centering
\includegraphics[width=5.2in,angle=0,trim=30 20 30 30]{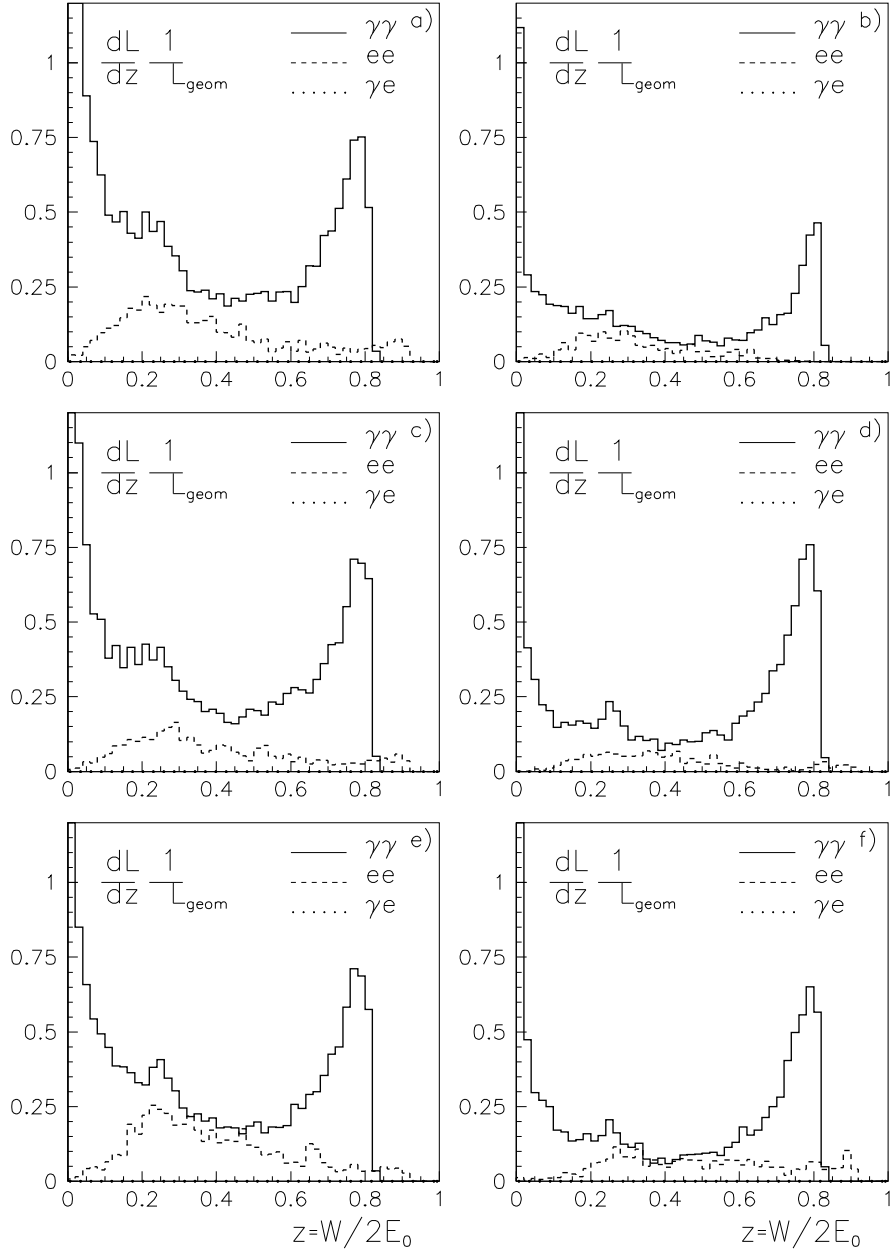}
\caption{Spectral \GG\ luminosity in the scheme with the \underline{vertical}
deflection for the beam parameters presented in 
Tables~\ref{table2},~\ref{table3} 
(Figs.~\ref{fig11}a), b)... correspond to TESLA(1), TESLA(2)...).
For additional comments see the text.}
\label{fig11}
\end{figure}

The optimization in this case consists of selection 
of CP--IP distance $b$ and a value of the magnetic field $B_e$. 
This choice was done using the following requirements and arguments:
\begin{enumerate}
\item[a)]  as soon as $\sigma_y < \sigma_x$ it is easier to deflect beams in 
the vertical direction;
\item[b)]  $b > 1.5\gamma\sigma_y$ (to suppress low energy luminosity);
\item[c)] the deflection at the IP $\Delta_y > 4\sigma_y$ (to avoid
collision with beamstrahlung photons);
\item[d)] the electromagnetic field at the IP should be below the threshold of
the coherent \EPEM\ creation ($\Upsilon < 1$).
\item[e)] b$>$1.5 cm (because the minimum distance of any material to the IP
should be larger than  1 cm, that is determined by 
synchrotron radiation from the
final quads (r$<$6 mm) and by low energy positrons which get kick
in the field of opposing beam and spiraling in the vacuum chamber
(see sect.5  for details).
\end{enumerate}
It turns out that it is possible to meet all these requirements using
$B_e \sim 0.5$ T. Such field can be produced by thin pulse magnet
surrounding the IP~ (see sect.4.8). 
The results on \GG\ collisions in the scheme with vertical deflection 
are shown
in Fig.~\ref{fig11} and summarized in Table~\ref{table3} (lane 2). 
One can see the following:
\begin{itemize}
\item at high invariant masses \LGG(defl.) $\sim$ \LGG(no
defl.); but the total \LGG(defl.) $\sim$ 0.2--0.3 \LGG(no defl.) and
one can expect smaller hadronic backgrounds (see sect.5);
\item $\LGE(z>0.65)\ll \LGG(z>0.65)$;
\item \LEE\ is negligibly small;
\item the disruption angles are less than about 5 mrad (10 mrad without
deflection).
\end{itemize}

\vspace{2mm}

\underline{\it \GE\ \it collisions}

\vspace{2mm}

  We have seen that in the scheme without
magnetic deflection, \GE\ collisions with a large luminosity can be
obtained simultaneously with \GG\ collisions. For some experiments 
it may be desirable to have pure \GE\ collisions (at least in the 
region of the high energy peak). This can be done using sweeping magnets.

  For optimization of \GE\ collisions with sweeping magnets we use the
following criteria: 

a) the spent electrons should be deflected in a
horizontal direction (larger beam size) because the ``main'' electron
beam is shifted during the collision with the opposing spent electron beam and
this shift should be smaller than the corresponding beam size. To
avoid direct collision of beams we have to provide larger magnetic
deflection than in the case of \GG\ collision that needs larger
distance between CP and IP; 

b) it should be checked that the process of coherent pair creation is
below threshold ($\Upsilon < 1$)  and broadening of the
luminosity spectrum due to beamstrahlung is rather small;

c) as before we assumed B=0.5 T (at larger fields (and smaller $b$)
the \LGE\ is somewhat larger, by 20-50\%). The distance $b$ between CP
and IP was chosen as some compromise between the point with highest \LGE\
in the high energy peak and the point where background is small enough. 

\begin{figure}[!hbtp] \centering
\includegraphics[width=5.2in,angle=0,trim=30 20 30 30]{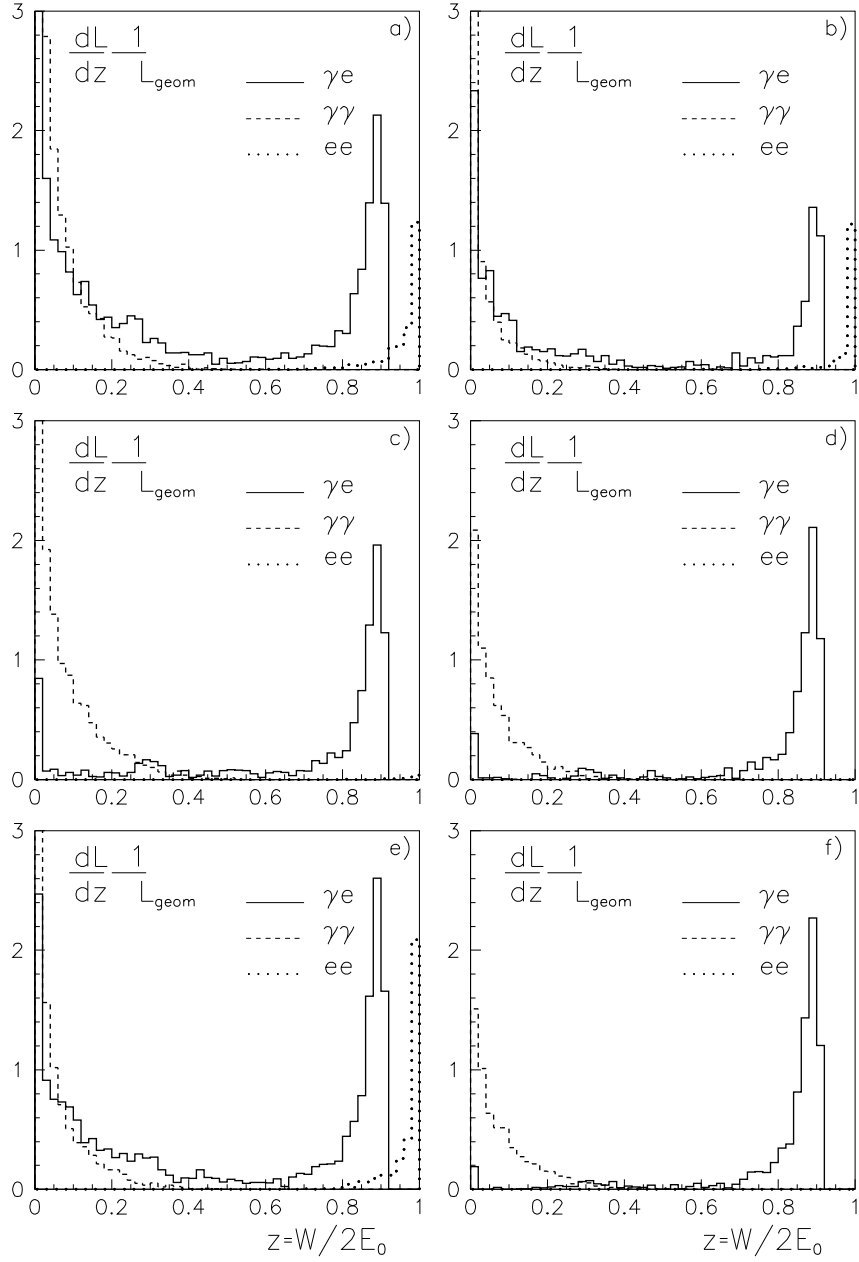}
\caption{Spectral \GE\ luminosity in the scheme with the
\underline{horizontal} deflection for beam parameters presented in
Tables~\ref{table2}, ~\ref{table3}.  (Figs.~\ref{fig12}a),
b)... correspond to TESLA(1), TESLA(2)...).  For additional comments
see the text.}  
\label{fig12} 
\end{figure}

The results on \GE\ luminosity (and \GG,~\GE\ backgrounds) with the
horizontal deflection are presented in Fig.~\ref{fig12}. and in 
Table~\ref{table3} (lane 3). 
Looking to these date one can make the following observations.

\begin{itemize} 
\item \GE\ luminosity spectrum is very
monochromatic, the full width at the half of the maximum is about 5\%.
\item \LGE($z>0.65) \sim (1.3 $--$ 5.2)10^{33}$ \CMS, 
approximately the same
as without deflection, but now it is more monochromatic and almost
without backgrounds.
\item In all cases there is unremovable low energy \GG\
luminosity. It is due to collisions of the beamstrahlung photons
(emitted by the ``main'' electron beam) with opposing high energy
Compton photons. One should also add collisions of virtual
(equivalent) photons with Compton photons which are not shown in our
figures.  
\item The low energy \GE\ and high energy $ee$ luminosity
in figs~\ref{fig12} a), b), e) (cases with large $\sigma_x$) are connected 
with the collision of tails of the ``main'' electron beam with the swept 
electron beam. Likely it will not cause problems for the analysis.
 Further increase of $b$ (length of the sweeping magnet) is not
desirable, because the transverse electron beam size at CP
$\sigma_x(b) =\sigma_x(0)(b/\beta_x)$ will be comparable with the laser
spot size and this leads to the decrease of the conversion coefficient. 
\end{itemize}

\begin{figure}[!hbtp]
\centering
\includegraphics[width=5.3in,angle=0,trim=30 20 30 30]{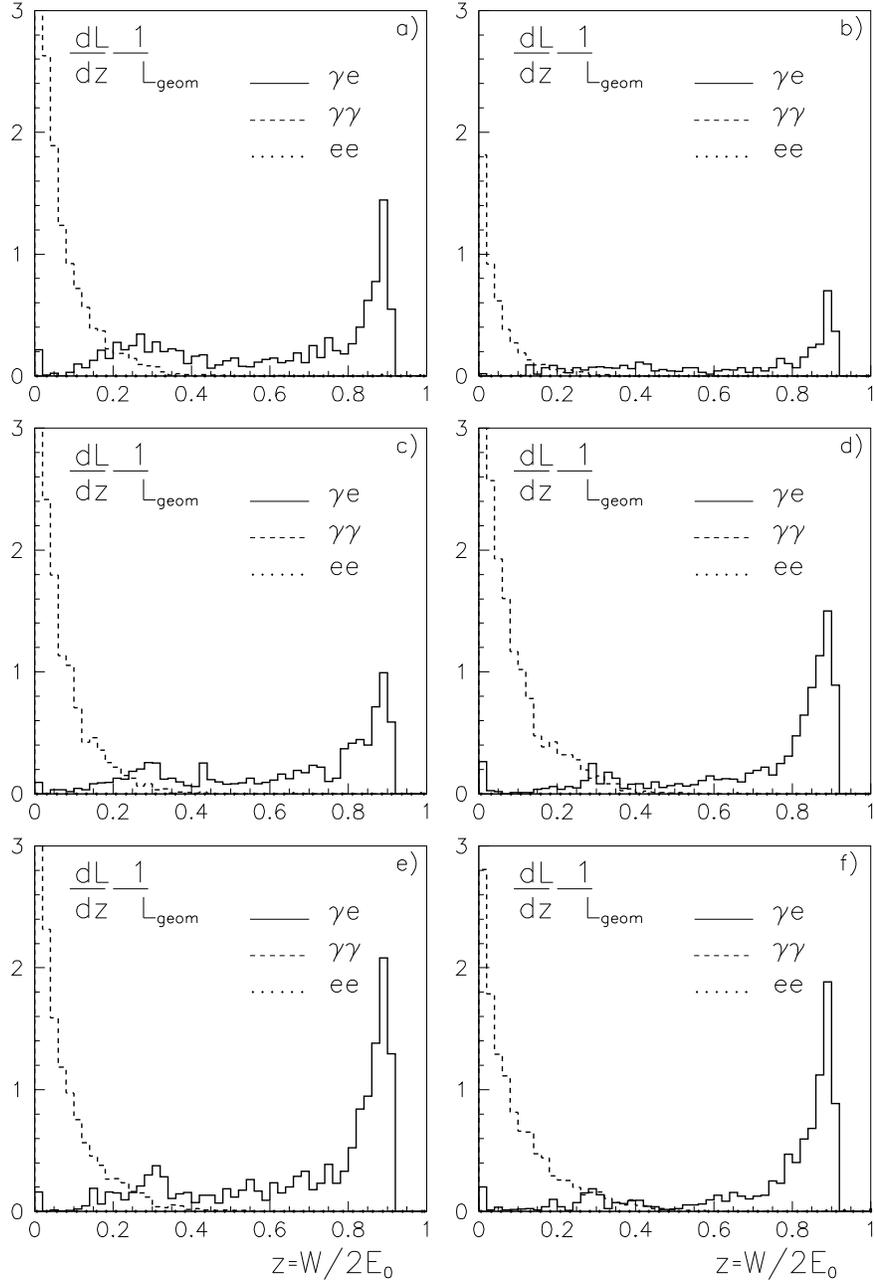}
\caption{Spectral \GE\ luminosity in the scheme with the same vertical 
deflection as for \GG\ collisions (fig.~\ref{fig11}) for beam parameters 
presented in Tables~\ref{table2},~\ref{table3}.
(Figs.~\ref{fig13}a), b)... correspond to TESLA(1), TESLA(2)...).
For additional comments see the text.}
\label{fig13}
\end{figure}

We have seen that \GE\ luminosity with optimized horizontal deflection
is large enough and has good quality. Unfortunately this requires change
of the sweeping magnet and some shift of optical elements. Therefore
it was interesting to check what are \GE\ luminosities when 
exactly the same deflection as for \GG\ case (vertical deflection) is used,
but with one laser switched off and the ``main'' electron beam is
somewhat shifted to collide with a high energy core of the
$\gamma$-beam. The results for this case are presented in Fig.~\ref{fig13} and
Table~\ref{table3} (lane 4). We see that the high energy peak is lower and
broader than with the horizontal deflection. This is mainly due to
beam displacement in the field of the opposing beam. It is of interest
that in some cases this displacement is much larger than $\sigma_y$ and
$b/\gamma$, but some luminosity at $z\sim z_{max}$ has survived. This
is because the displacement during beam collisions grows quadratically with
the passed distance and on the first part of collision length it is
relatively small.

\vspace{2mm}

\subsubsection{ Ultimate luminosities}

\vspace{2mm}

\begin{figure}[!hbtp]
\centering
\includegraphics[width=5.3in,angle=0,trim=30 20 30 30]{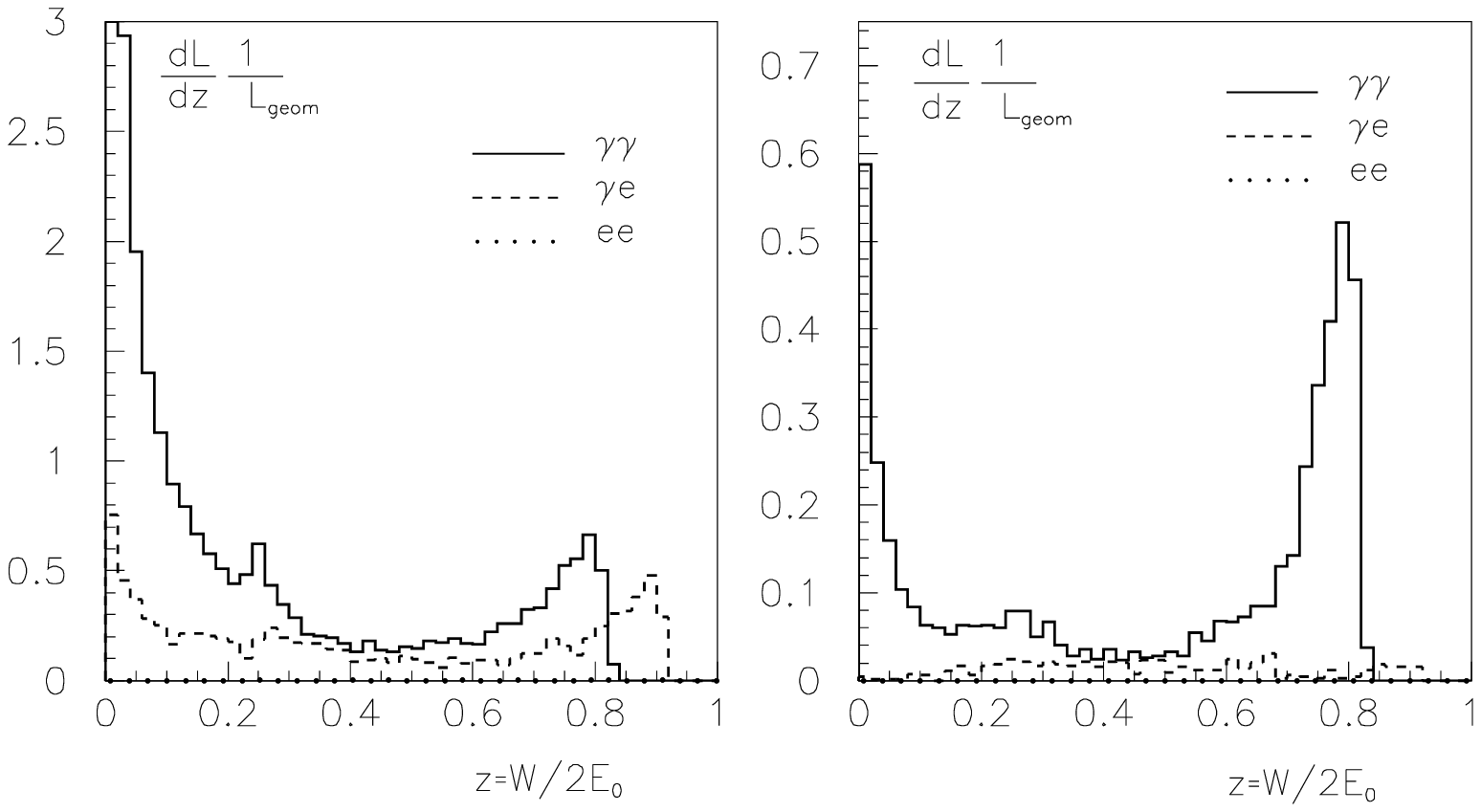}
\vspace*{-1cm}
\caption{Luminosity spectra for the ``super TESLA(4)''
parameters (see the text). Left figure - without the deflection;
right - \GG\ collisions with the magnetic deflection.} 
\label{fig14}
\end{figure}

From the results presented in Table~\ref{table3} follows 
that \LGG$(z>0.65)$ in
the considered range of parameters is approximately proportional
to the $L_{geom}$ and the case TESLA(4) has
maximum luminosity. What further improvement is possible providing
the problem of low horizontal emittances is solved (laser cooling, for
example)? Let us consider the case of TESLA(4) but with 5 times smaller
emittances: $\epsilon_{xn}=\epsilon_{yn}=2\times10^{-7}$
~m~rad. The results are presented below and in Fig.~\ref{fig14}.

{\centering{``Super TESLA(4)''}

$N = 3.63 \times 10^{10}$, $\sigma_z = 0.5$~mm, $2E = 500$~GeV, $f = 5.65$~kHz,
$\epsilon_{nx} = \epsilon_{ny} = 0.2 \times 10^{-6}$~m$\times $~rad, 
$\beta_x = \beta_y = 0.5$~mm, $\sigma_x=\sigma_y=14$~nm, 
$L_{geom}=2 \times 10^{35}$~cm$^{-2}$~c$^{-1}$

\underline{no deflection: $b = \gamma \sigma_y = 0.7$~cm},

$L_{\GG}=1.15 \times 10^{35}$ ,
$L_{\GG} (z>0.65) = 1.5 \times 10^{34}$~cm$^{-2}$~c$^{-1}$, \\
$L_{\GE}=3.6 \times 10^{34}$,
$L_{\GE} (z>0.65) = 1.2 \times 10^{34}$~cm$^{-2}$~c$^{-1}$,  \\
$L_{\EE}=1.5 \times 10^{32}$,
$L_{\EE} (z>0.65) = 8 \times 10^{31}$~cm$^{-2}$~c$^{-1}$,   \\
$L_{\EPEM}=6 \times 10^{32}$,
$L_{\EPEM} (z>0.65) = 1.4 \times 10^{32}$~cm$^{-2}$~c$^{-1}$, \\
$L_{\GP}=5.6 \times 10^{32}$,
$L_{\GP} (z>0.65) = 1.2 \times 10^{31}$~cm$^{-2}$~c$^{-1}$,  
\vspace{3mm}
  
\underline{with magnetic deflection: $b = 1.5$~cm, $B = 0.5$~T},

$L_{\GG}=2 \times 10^{34}$ ,
$L_{\GG} (z>0.65) = 1 \times 10^{34}$~cm$^{-2}$~c$^{-1}$, \\
$L_{\GE}=2.5 \times 10^{33}$,
$L_{\GE} (z>0.65) = 6 \times 10^{32}$~cm$^{-2}$~c$^{-1}$,  \\
$L_{\EE}=5 \times 10^{30}$ ~cm$^{-2}$~c$^{-1}$

 }

   Results are impressive: \LGG $(z>0.65)\geq 10^{34}$ \CMS ~in both case
with and without deflection. Let us note one interesting fact. The
beams in this example have transverse sizes $\sigma_x=\sigma_y=14$~nm
and  the parameter $\Upsilon \sim 3$, higher than
the threshold of the coherent pair creation (see sect.4.2). In collisions
without deflection we could expect very high conversion probability of
the high energy photons to \EPEM\ pairs. But simulation shows that
there is  no such problem, though there are some small \EPEM,~\GP\
luminosities. An explanation for this is the following \cite{TELSH}: due to
repulsion beams are separated during the collision at a rather large
distance and their field on the beam axis (which influence on high
energy photons) is below the critical value $\Upsilon \sim 1$. 

  From this picture an incredible conclusion follows:  
 one can use for \GG\ collisions the electron beams even with
infinitely small transverse sizes (with $N,\;\sigma_z,\;E_0$ considered
above)! The \GG\ luminosity will be determined by the photon spot size
which is equal to $b/\gamma$. In this case~\cite{TELSH}
\begin{equation}
\LGG(z>0.65) \sim 0.35\frac{k^2N^2f}{4\pi(b/\gamma)^2} \sim 
 k^2\left(\frac{N}{10^{10}}\right)^2 \frac{E^2[\TEV\ ]
f[{\rm kHz}]}{b^2[\MM\ ]} \times 10^{36} 
\end{equation} 
 For TESLA with $b=3$~mm, $k^2=0.4$ we
get $\LGG(z>0.65)\sim 2\times10^{35}\;\CMS$, providing 
 $\sigma_x, \sigma_y \leq b/2\gamma = 3$ \NM. The simulation confirms
this result.
The last example may sound too fantastic, but $\LGG(z>0.65)\sim
10^{34}$~\CMS\  demonstrated before (with and without magnetic deflection) is
a good goal for ``upgraded'' TESLA (SBLC).

\begin{table}[!hbtp]
\caption{Parameters of the \GG, \GE\ collider. Corresponding 
parameters of electron beams are given in table 2. 
The cases of ultimate luminosities (sect. 4.5.3) are not included.}
\begin{center}
\begin{tabular}{c c c c c c c} \hline
 &\scriptsize TESLA(1) &\scriptsize TESLA(2) &\scriptsize TESLA(3) &
  \scriptsize TESLA(4) &\scriptsize SBLC(1) &\scriptsize  SBLC(2) 
                                                   \\ \hline \hline 
\multicolumn{7}{c}{$\GG$, no deflection, $b=1.5 \gamma \sigma_y$,
see also Fig.~\ref{fig10} } \\ \hline
b, cm & 1.17 & 0.41 & 1.17 & 2.35 & 1.04 & 1.28 \\
$\LGG, 10^{33}$ & 13.2 & 11. & 31. & 34. & 6.3 & 21. \\
$\LGG (z>0.65)$ & 1.2 & 1.5 & 2.7 & 3.3 & 0.7 & 2.7 \\
$\LGE, 10^{33}$ & 8.7 & 7 & 17 & 22 & 5.9 & 16 \\
$\LGE (z>0.65)$ & 2 & 2 & 4 & 5.5 & 1.6 & 5 \\
$\LEE, 10^{33}$ & 0.86 & 0.6 & 0.72 & 1 & 1.05 & 1.15 \\
$\LEE (z>0.65)$ & 0.55 & 0.39 & 0.48 & 0.74 & 0.69 & 0.78 \\
$\theta_x(\theta_y)_{max}, mrad$ & 6.5(7.5) & 5.5(6) & 6.5(9)& 
 7.5(8) & 4.5(6) & 6(6) \\ \hline
\multicolumn{7}{c}{$\GG$, with vertical deflection, $B=0.5$~T,
see also Fig.~\ref{fig11} } \\ \hline
b, cm & 1.5 & 1.5 & 1.5 & 2.35 & 1.5 & 1.53 \\
$\LGG, 10^{33}$ & 4.9 & 2.2 & 10.2 & 8.2 & 2.3 & 6.2 \\
$\LGG (z>0.65)$ & 1.1 & 0.67 & 2.5 & 3.3 & 0.6 & 2.3 \\
$\LGE, 10^{33}$ & 1.0 & 0.47 & 1.65 & 1.0 & 0.67 & 1.6 \\
$\LGE (z>0.65)$ & 0.16 & 0.01 & 0.25 & 0.14 & 0.1 & 0.45 \\
$\theta_x(\theta_y)_{max}, mrad$ & 2.5(5) & 2(4) & 2.0(5.5)&
 0.5(3.5) & 2(3) & 1(2.3) \\ \hline
\multicolumn{7}{c}{$\GE$, with horizontal deflection, $B=0.5$~T,
see also Fig.~\ref{fig12} } \\ \hline
b, cm & 3.5 & 3 & 3.5 & 4.5 & 3.0 & 2.5 \\
$\LGE, 10^{33}$ & 5.6 & 3.5 & 5.6 & 6.0 & 2.8 & 5.3 \\
$\LGE (z>0.65)$ & 2 & 1.3 & 4.0 & 5.2 & 1.3 & 4.8 \\
$\LGG, 10^{33}$ & 5.8 & 2.4 & 7.7 & 5.3 & 1.7 & 3.9 \\
$\LGG (z>0.65)$ & $<0.001$ & $<0.001$ & $<0.001$  & $<0.001$ & $<0.001$ &
 $<0.001$ \\
$\theta_x(\theta_y)_{max}, mrad$ & 2.5(2.5) & 2(2) & 2(2.5)&
 2.5(2.5) & 2(2) & 2(2) \\ \hline
\multicolumn{7}{c}{$\GE$, with vertical deflection, optimized for $\GG$,
 $B=0.5$~T, see also Fig.~\ref{fig13} } \\ \hline
b, cm & 1.5 & 1.5 & 1.5 & 2.35 & 1.5 & 1.53 \\
$\LGE, 10^{33}$ & 2.5 & 1.2 & 4.9 & 7.5 & 2. & 6. \\
$\LGE (z>0.65)$ & 1.35 & 0.72 & 2.9 & 5.5 & 1.3 & 4.6 \\
$\LGG, 10^{33}$ & 3.7 & 1.5 & 8.1 & 13 & 1.9 & 7.8 \\
$\LGG (z>0.65)$ & $<0.001$ & $<0.001$ & $<0.001$  & $<0.001$ & $<0.001$ &
 $<0.001$ \\
$\theta_x(\theta_y)_{max}, mrad$ & 3(5.5) & 2(3.5) & 2.5(5.5)&
 2(6) & 1(3) & 2(2.5) \\ \hline

\end{tabular}
\end{center}
\label{table3}
\end{table}

\subsection{Summary table of \GG,\GE\ luminosities}

 The results on \GG, \GE\ luminosities collected on Table 3 and corresponding
figures can be summarized as follows. 
\vspace{2mm}

\centerline{\large \underline{\GG\ -luminosities}}
\vspace{2mm}

  High energy \GG\ luminosity $L_{\GG}(z\ge0.65)$\ is approximately the 
same in both collision schemes (with and without the deflection) and 
account for

\begin{center}
$L_{\GG}(z\ge0.65)\sim10^{33}$ \CMS\ for ``nominal'' (TESLA(1), SBLC(1))
parameters. \\
$L_{\GG}(z\ge0.65)\sim (1-3)\times10^{33}$ \CMS\ for beam parameters 
presented in Table 2 and in the top part of Table 3.
\end{center}

   In the case of progress in obtaining electron beams with 
lower horizontal emittance one can get

\begin{center}   
$L_{\GG}(z\ge0.65)\sim10^{34}-10^{35}$ \CMS\ !
\end{center}

  The peak luminosity  is also an important characteristic. For all considered 
cases it is approximately

\begin{equation}
\frac{dL_{\GG}}{dz} z_{max} \sim 7L_{\GG}(z>0.65).
\end{equation}

  The ratio $L_{\GG}(\textrm{total})/L_{\GG}(z\ge0.65)$ is 
a less definite parameter, which depends on the distance between
the CP and IP and on the value of the sweeping field (in the case
with deflection). For considered variants, it  changes 
between 2--4 and 8--11 for the schemes with and without deflection
respectively. The hadronic background in the second case will be
larger by a factor of 2 (see sect.5).
\vspace{2mm}

\centerline{\large \underline{\GE\ -luminosities}}
\vspace{2mm}

  $L_{\GG}(z\ge0.65)\sim(1.5-5)\times10^{33}$ \CMS\ in both 
collision schemes. In the case with magnetic deflection the 
luminosity spectrum is quite monochromatic (FWHM$\sim5$\%).

 A priori it is clear that pure \GG\ and \GE\ collisions
with lower background in the scheme with deflection are simpler
for  analysis, but the scheme without deflection 
also has some positive features: no sweeping magnets, simultaneous
\GG, \GE\ collisions, higher luminosity at low and intermediate 
invariant masses (they could be even higher with lower b).
Further studies should show how serious  the problems with analysis 
at  these conditions are.

\subsection{ Monitoring and measurement of \GG, \GE\ luminosities}

\IND\ A system produced in a \GG\ collision is characterized by its
invariant mass W$_{\GG} = \sqrt{4\omega_1\omega_2}$ and rapidity
$\eta=0.5\ln(\omega_1/\omega_2)$. We should have a method to measure
1) $d^2$L/$d$W$d\eta$ and 2)$\lambda_{\gamma 1}\lambda_{\gamma 2}$
or, in other words, $d$L$_0$/$d$W$d\eta$ and $d$L$_2$/$d$W$d\eta$
(0,2---total helicity of the system). This can be measured using the
process \GG\ $\to$ \EPEM\ ($\mu^+\mu^-$) \cite{TEL93,MILLER}. 

For this process $\sigma_0/\sigma_2\sim m^2/s$ (excluding
the region of small angles), $s=4E_0^2$.  Therefore, the measurement of this
process will give us $d$L$_2$/$d$z$d\eta$.  How to measure
$d$L$_0$/$d$z$d\eta$? This can be done by inversion of the
helicity($\lambda_{\gamma}$) of the one photon beam by means of
changing simultaneously signs of helicities of the laser beam used for
${\rm e}\rightarrow\gamma$ conversion and that of the electron beam
(in photo injector).  In this case the spectrum of scattered photons
is not changed while the product $\lambda_{\gamma 1}\lambda_{\gamma
2}$ changes its sign. In other words, what was before L$_0$ is now
L$_2$, which we can measure.  The cross section for this process
$\sigma(|\cos\Theta|<0.9)\approx 10^{-36}$/s[TeV$^2$], cm$^2$.  This
process is very easy to select due to a ze\-ro co\-pla\-na\-ri\-ty
angle.

  Other processes with large cross sections which can be used for
the luminosity measurement are $\GG\ \to \EPEM\ \EPEM\ $ \cite{GKST83} and
\GG\ $\to W^+W^-$ \cite{YASUI}. The first process has the total cross
section of $6.5\times10^{-30}\;$ cm$^2$, the second one $8\times10^{-35}\;
$cm$^2$. Unfortunalely, both these processes practically do not depend  on
the polarization and the first one is difficult for detection.
  
   For the absolute \GE\ luminosity measurement one can use the
process of Compton scattering which is strongly polarization
dependent.

  For luminosity tuning one can use beam-beam deflection and
background processes such as \EPEM\ and hadron production. 
In the scheme without sweeping magnets the beam deflection method is
sufficient. It is more difficult to control \GG\ collisions in the
scheme with the magnetic deflection. Besides the process $\GG\ \to hadrons$,
there are other possibilities which were not yet studied,  such as
beam-beam deflection, \EPEM\ production by a high energy photon in a collision
with swept electrons (at large impact distances) or \EPEM\ pair production in a
collision with synchrotron photons from the sweeping magnet. Using the
hadronic production as a ``firm'' signal of the \GG\ luminosity (hundreds
events per train collision) one should continuously measure other
collision characteristics  enumerated above and use fast feedback
(inside bunch train duration) to stabilize conditions when the hadron yield is
maximum~\cite{TEL}.

\subsection{Sweeping magnet}

  Only few remarks on the sweeping magnet.  It can be a thin one loop
pulse magnet \cite{SIL}. Considerations show that for $B=0.5\;T$ there
are now visible problems. Some arguments:
 
\begin{itemize}
\item This
field produces a 1 atm pressure, for a few cm magnet this makes no
problems.
\item For 2 mm Al coil thickness and 3 cm magnet length the
total average power is about 100 W for TESLA time structure and
even less for SBLC. 
\item The thickness of the skin layer for Al alloy for
$\nu \sim$ 0.25 kHz is about 5 mm.  Attenuation of the field by about 0.5
mm thick Al vacuum pipe will be small (can be compensated), may be
some cooling of the vacuum chamber will be necessary. For SBLC (with 2
$\mu$s bunch train length) the duration of current pulses should be
optimized taking into account this effect.  With a dielectric vacuum
pipe there will be no such problem. 
\item One loop pulse magnet contains no dielectric material and can be
 put (if necessary) inside the vacuum pipe.
\item Contribution of multiple scattering at the magnet to the 
impact parameter resolution is about $\sigma = 30/P[\GEV],\;\MKM\ $ for
$\vartheta =\pi/2$, which is acceptable. 
\end{itemize}

\section {Backgrounds}

   What are backgrounds at \GG,\GE\ collider? Are they larger or 
smaller than in \EPEM\ collisions? Let us try to answer these questions.

  At \GG,\GE\ colliders there are following sources of backgrounds:

a) particles with large disruption angles hitting the final quads. 
The sources are multiple Compton scattering, hard beamstrahlung,
Bremsstrahlung (in \EE ).

b) \EPEM\ pairs created in the processes of $\EE  \to  \EE \EPEM$~
(Landau-Lifshitz, LL), $\GE \to \textrm{e} \EPEM$~(Bethe-Heitler, BH),
$\GG \to \EPEM$~(Breit-Wheeler, BW). This is the main source
of low energy particles, which can cause problems in the vertex detector.

c) $\GG \to hadrons$, $\GG \to \EPEM \EPEM$\ --- the processes with largest
cross-sections in \GG\ collisions.

  We have already discussed the \underline{item a)} before 
and found that the minimum energy of these particles is about
 2\% of $E_0$, the maximum disruption angle is below 10 mrad and they 
can be removed using the crab crossing scheme.

  The Bremsstrahlung process $\;\EPEM \to \EE \gamma\;$ is suppressed
at photon colliders (due to beam repulsion or magnetic deflection).
In the scheme without deflection 
$L_{\EE}/ \textrm{crossing}
\sim 10^{33}/10^4 = 10^{29}\;$ \CMS\ per beam crossing.
The low energy electrons after hard bremsstrahlung can have an energy 
below 2\% $E_0$ and therefore hit the quads. The total energy of
these particles for $L_{ee}$ given above
is about $1.5 E_0$~ per bunch crossing, which is small.

   \underline {The item b)} for \GE, \GG\ colliders is even less 
important than for \EPEM\ colliders because one of the main sources (LL)
is almost absent. Nevertheless, we will consider here  main
characteristics of \EPEM\ pairs which are important for designing
sweeping magnets (aperture) and  vacuum chamber near the IP.

\begin{figure}[!hbt]
\centering
\includegraphics[width=4.5in,angle=0,trim=30 0 30 10]{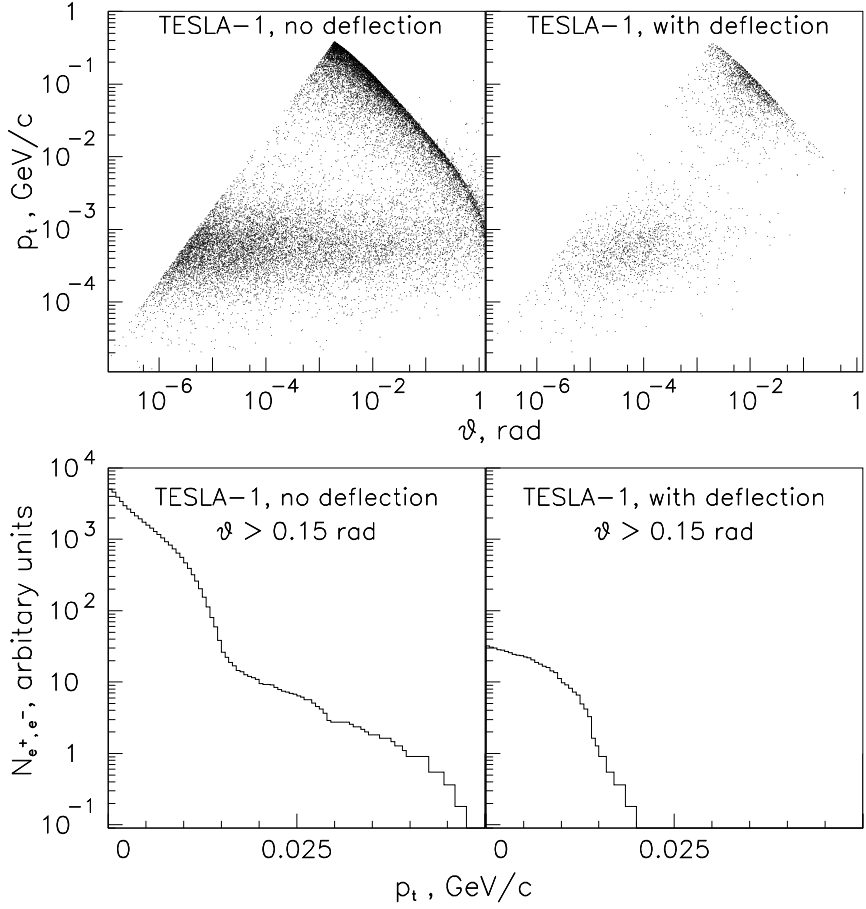}
\caption{Upper figures: 
scatter plots of transverse momentum vs azimuthal angle,
lower figures: $p_{\bot}$ distribution of \EPEM\ with  
$\theta\ge0.15$~rad.}
\label{fig15}
\end{figure}
   
   Most of $e^-$ and $e^+$ produced in LL, BH, BW processes travel
in a forward direction, but due to the kick in the field of the
opposing electron beam they get much larger angles and can cause 
problems in the detector. We have simulated these processes for
two situations: 1) TESLA(1) without deflection (see Fig.~\ref{fig10}a)
and Table~\ref{table3}), 2) TESLA(1)  with  magnetic deflection
(see Fig.~\ref{fig11}a).
The total number of \EPEM\ pairs produced in these cases per one 
bunch collision is 46000 and 7000 respectively.
The distributions are presented in Fig.~\ref{fig15}. In the plot
$p_{\bot}$ vs $\vartheta$ we see two regions of points concentration.
The region with large $p_{\bot}$ corresponds to electrons 
deflected off the opposing beam (it was assumed in the simulation that
$e^+$ have no kick). In the second case (with deflection) the 
backgrounds are much smaller.

Any materials near the IP should be
located beyond the zone occupied by kicked particles. The background
due to the particles with large initial  angles is rather small.
The shape of the zone occupied by the kicked electrons is 
described by the formula  ~\cite{BATTEL}
\begin{equation}
r^2_{max} \simeq \frac{25Ne}{\sigma_zB}z \sim 0.12\frac{N}{10^{10}}
\frac{z\textrm{[cm]}}{\sigma_z\textrm{[mm]}B\textrm{[T]}}
\end{equation} 

\noindent where $r$ is the radius of the envelope at a distance 
$z$ from the IP, $B$ is longitudinal detector field.
For example, for TESLA ($N=3.63\times10^{10}, \sigma_z=0.5$~mm,
$B=3$~T) $r=0.55 \sqrt{z\textrm{[cm]}}$, cm. This simple formula
can be used for the choice  of a vertex detector radius and
aperture of the sweeping magnet. At $z=b\simeq3$~cm 
(sweeping magnet) we have $r \simeq 0.95$, cm.
For the cylindrical vertex detector with $l=\pm15$~cm $r=2.15$~cm.

  Note that the sweeping magnet (with transverse field) gives to
all particles $p_{\bot}=e B_e l/c$, where $l$ is CP--IP distance.
For $B_e=0.5$~T and $l=2$~cm $p_{\bot}=3$~MeV/c. The spiral radius for 
these electrons in 3 T field is 0.33 cm. This is not dangerous, but
larger  $B_e$ could make problems. This is very important point
restricting the value of field in the sweeping magnet. 

  The total energy of all \EPEM\ produced in the considered
processes per one beam collision is $1.6\times10^6$~GeV and
$1.7\times10^5$~GeV respectively for the cases TESLA(1, without
deflection) and  TESLA(1, with deflection). But most of these particles
escape the detector inside a 10 mrad cone without interaction,
and only particles with $\vartheta>10$~mrad and $p\lesssim 1$~GeV
(due to crab crossing in the solenoidal field) will hit the quads.
The total energy of these particles is much smaller: $2\times10^4$ and
$0.5\times10^4$~GeV for the two cases respectively. The conical mask
(the same as for \EPEM\ collision) around the beam can protect
the detector from the low energy backscattered particles from
electromagnetic showers.

\begin{figure}[!hbt]
\centering
\includegraphics[width=4.5in,angle=0,trim=30 -5 30 30]{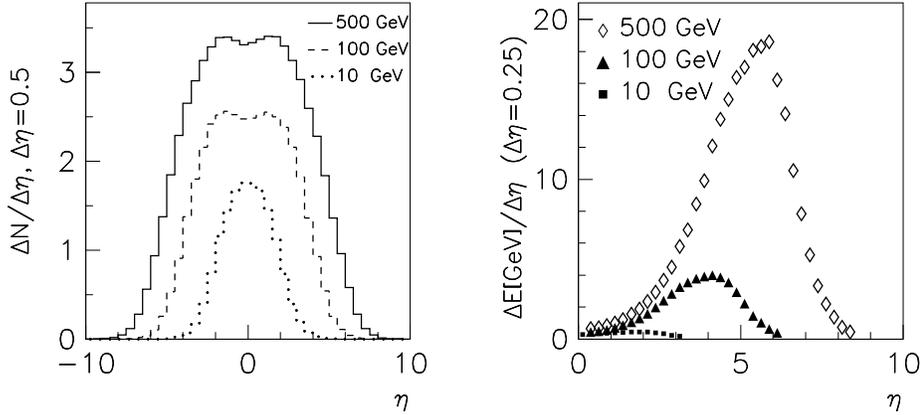}
\caption{Distribution of particle (left) and energy flow (right)
on pseudorapidity in $\GG \to hadrons$\ events (photons have
equal energies.}
\label{fig16}
\end{figure}

\begin{figure}[!hbt]
\centering
\includegraphics[width=1.2in,angle=0,trim=200 100 200 120]{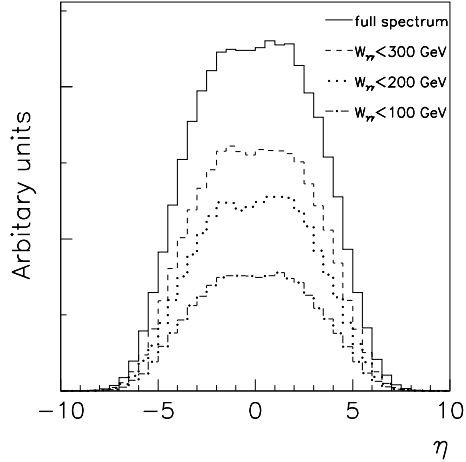}
\caption{Distribution of particles 
on pseudorapidity for different ranges of \GG\ 
invariant mass for the TESLA(1) case without  deflection (see 
Fig.~\ref{fig10}a).}
\label{fig17}
\end{figure}

\begin{figure}[!hbt]
\centering
\includegraphics[width=4.in,angle=0,trim=30 0 30 30]{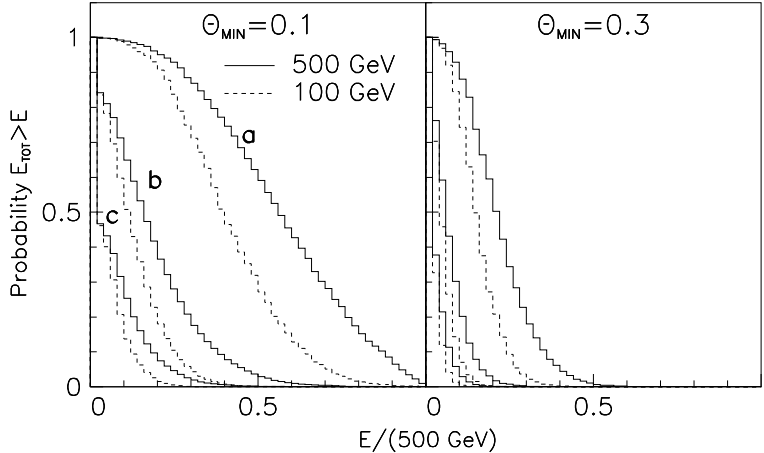}
\caption{Probability of energy deposition in the detector above some
value ($E/500$ GeV) due to the process $\GG \to hadrons$. 
The polar angle acceptance is
$\vartheta > 0.1$~rad (left plot) and  $\vartheta > 0.3$~rad (right plot).
Curves a), b), c) correspond to 7, 2 and 0.7 hadronic events on the average
per beam collision respectively. The collision energy is 500 GeV (solid line)
and 100 GeV (dashed line), photons have equal energies.}
\label{fig18}
\end{figure}

\begin{figure}[!htb]
\centering
\includegraphics[width=3.9in,angle=0,trim=30 40 30 40]{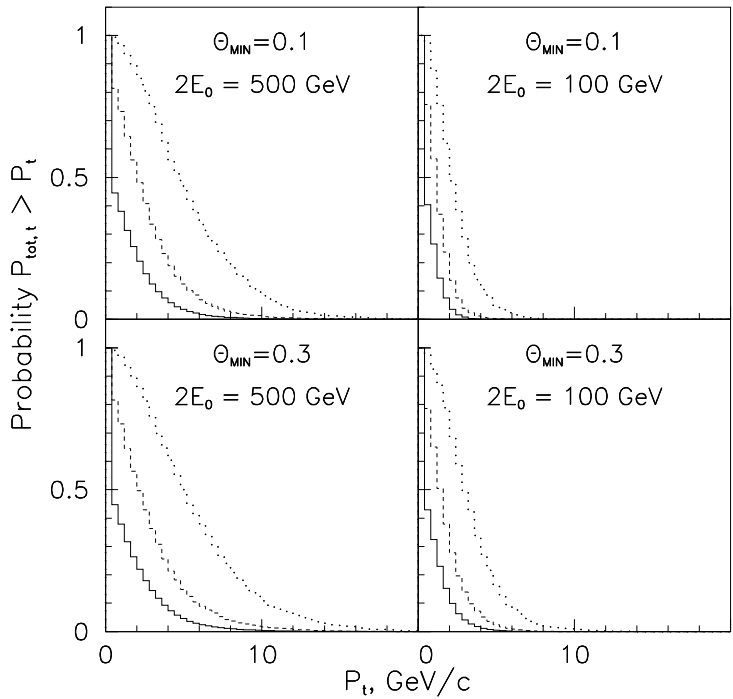}
\caption{Probability to find an unbalanced transverse 
momentum above some $p_t$. Dotted, dashed and solid curves 
correspond to 7, 2, 0.7  $\GG \to hadrons$ events on the average per 
beam collision. The polar angle acceptance is
$\vartheta > 0.1$~rad (upper plots) and  $\vartheta > 0.3$~rad (lower plots).
The collision energy is 500 GeV (left plots) and 100 GeV (right plots),
photons have equal energies.}
\label{fig19}
\end{figure}

\begin{figure}[!htb]
\centering
\includegraphics[width=4in,angle=0,trim=30 20 30 20,clip]{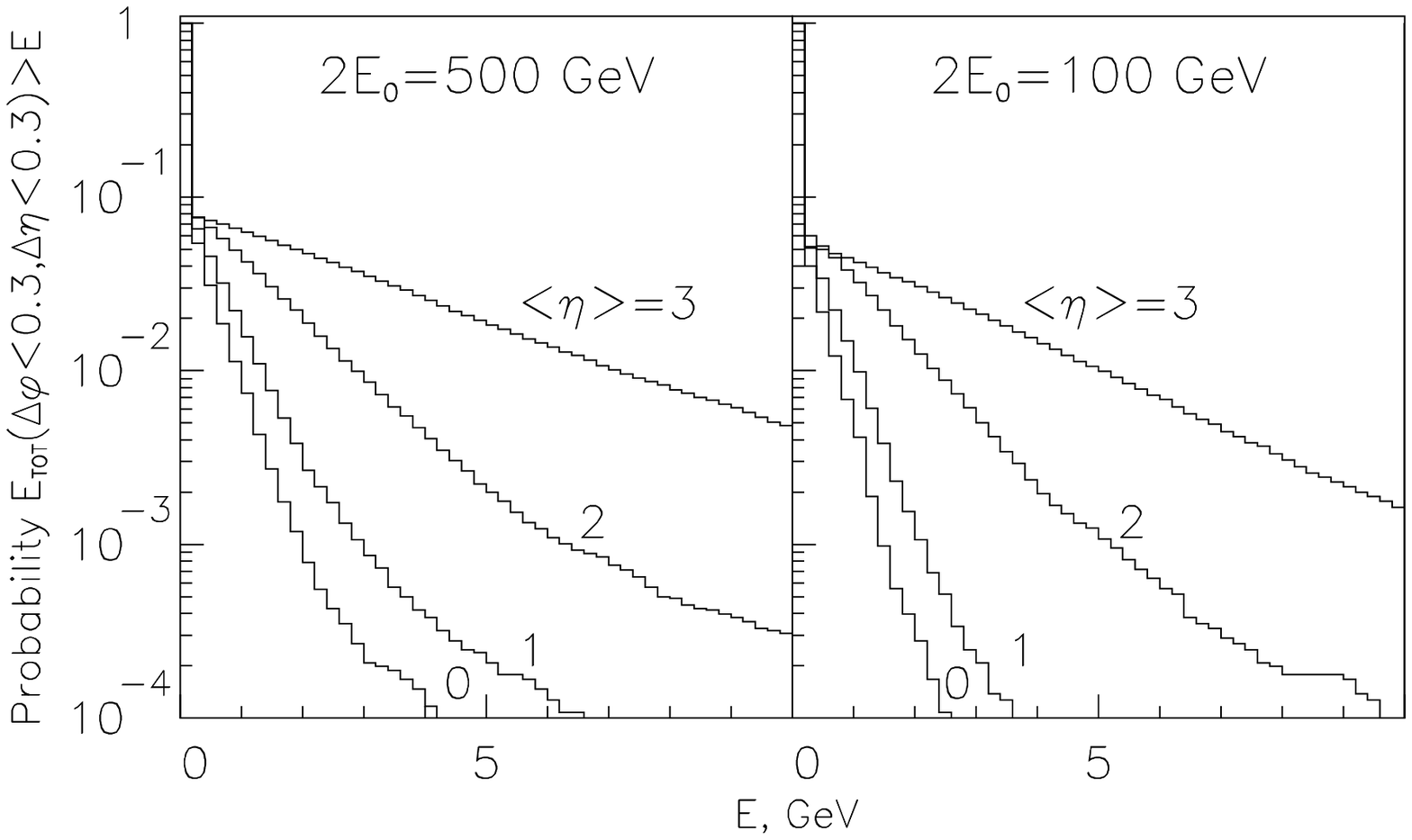}
\caption{Probability to have the energy flow into
$\Delta\phi \times \Delta\eta=0.3\times0.3$\ cell above
some threshold (abscissa value) for 4 pseudorapidity points: 
$\eta=$0, 1, 2, 3. The energy of symmetrical \GG\ is 500 GeV (left)
and 100 GeV (right).}
\label{fig20}
\end{figure}

  \underline {The item c)} $\GG \to hadrons$ is the most
specific background for a photon collider. The cross section 
of this process is about $400-600$~nb at $2E=10-500$~GeV.
The \GG\ luminosity is $2\times10^{29}-5\times10^{30}$~\CMS\ 
per bunch crossing (see Table~\ref{table3}), which leads to
$0.1-2.5$ events per beam collision. We considered also 
the cases of ``superluminosities''
which have even more events per crossing. What does
it mean for experiment and where is the limit? To understand this we have
performed  simulation using the PYTHIA code 5.720 ~\cite{SCHULER}.
Fig.~\ref{fig16} shows the distributions of particles and energy
on pseudorapidity ($\eta=-{\ln {\tan(\vartheta/2)}}$) in one
$\GG \to hadrons$ event at $2E=10, 100$ and 500 GeV. We see that
each 500 GeV hadronic event gives on the average 25 particles 
(neutral $+$ charged) in the range of 
$-2\le \eta \le2$ ($\vartheta\ge0.27$~rad) with the total energy about 15 GeV.
The average momentum of particles is about 0.4 GeV.   
Note that the flux of particles at large angles ($\eta=0$)
from 10 GeV event is only twice smaller than that from a 
500 GeV \GG\ collision.

 In this respect it is of interest to check what background gives different 
parts of \GG\ luminosity spectra for the scheme without deflection.
Fig.~\ref{fig17} also shows the distribution of particles on 
pseudorapidity for the TESLA(1) case. We see that the suppression of
\GG\ luminosity in the
region of $W_{\GG}\le200$~GeV using the magnetic deflection 
decreases the hadronic backgrounds by a factor of two. 
     
   The probability of energy deposition in the detector above some value
($E/500$ GeV) is shown in Fig.~\ref{fig18}. In the left figure
the minimum angle of the detector is 0.1 rad, on the right one
0.3 rad. Curves a), b), c) correspond to 7, 2 and 0.7 hadronic events on 
the average per collision; solid curves are for $2E=500$ GeV,
dashed are for 100 GeV. For 7 events per collision and $\vartheta_{min}=0.1$
the energy deposition in the detector with 50\% probability exceeds
55\% of $2E_0$. But this energy is produced by about 300 rather soft 
particles. This smooth background can be subtracted.
More important are the fluctuations of background. 
Below there are some of such characteristics.

  In many experiments the important characteristic is a missing
transverse momentum. The probability to find an unbalanced transverse 
momentum above some $p_t$ is shown in Fig.~\ref{fig19} 
for $\vartheta_{min}=0.1$
and 0.3, for 500 and 100 GeV \GG\ collisions. Again 3 curves
in each figure correspond to 7, 2, 0.7 hadronic events
on the average per collision. The results are interesting. Comparing
the curves at $\vartheta_{min}=0.1$ and  $\vartheta_{min}=0.3$ we see
that the difference is small, which is surprising because in the 
second case the energy deposition in the detector is by a factor of 2.6 
smaller. The explanation is the following: in the second case 
the detector measures smaller part of the total energy and the 
fluctuations are larger (at $\vartheta_{min}=0$ all particles are detected
and in a perfect detector unbalanced $p_{\bot}=0$). Let us look at the numbers.
Even with 7 events (500 GeV) per collision the probability to get
unbalanced  $p_{\bot}\ge 10$~GeV is only about 10\%, which is almost
acceptable.

   While  calculating $p_{\bot}$, we summed all energy depositions in the 
detector, but ``interesting'' events usually have high energetic particles
or jets. Let us check what  the probability that the hadronic 
background adds some additional energy to a jet. This information
is presented in Fig.~\ref{fig20}.
We have selected a cell $\Delta\varphi\le 0.3$,
$\Delta\eta\le0.3$, which  corresponds to a characteristic jet transverse
size, and calculated the probability of energy deposition in this region above 
some energy $E$. The curves correspond to one hadronic event on the average per
bunch collision. For other levels of background the probability should be 
multiplied by the average number of hadronic events per collision. A typical
energy resolution for 100 GeV jet is about 4 GeV. The probability
to have such energy deposition at $\eta=0$ and 10 events per collision
is less than 0.1\%, at $\eta=2$~($\vartheta=0.27$~rad) it is 4\%, which is
acceptable. Note that the average number of background particles
in  $\Delta\varphi\ \times \Delta\eta = 0.3\times0.3$~ cell 
for a 500 GeV \GG\ collision is $\sim0.1\times n_{ev/coll}$, so for
$n\le10$ the energy depostion (which we used) and its r.m.s. fluctuations
are close to each other.

   So, we can conclude that even 10 hadronic events in a 500 GeV \GG\
collision are acceptable for experiments. In the central part 
of the detector ($\eta\le1$~,$\Delta\Omega /4\pi = 0.76$)
even 50 events per collision lead to only 5\% probability 
that background shifts the jet energy by one r.m.s. energy resolution.
This condition corresponds at TESLA to $L_{\GG}=5\times10^{35}$~\CMS.

\section{Optics in the interaction region.}

\begin{figure}[thb]
\centering
\includegraphics[width=5.3in, angle=0, trim=0 0 0 0]{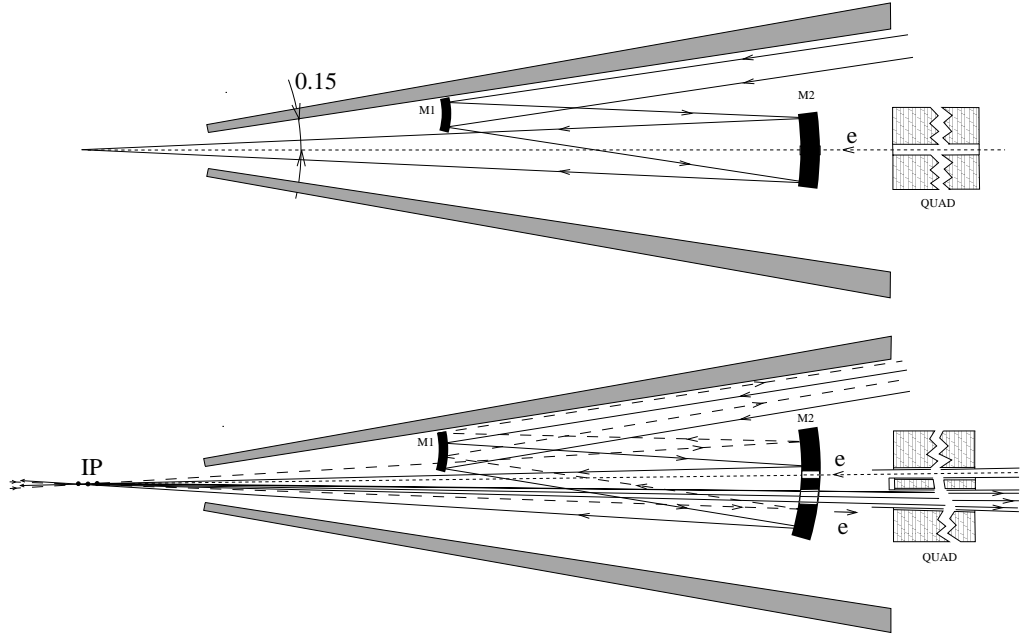}
\caption{Layout of laser optics near the IP; upper - side view,
down - top view, dashed lines - exit path of light coming from the
left through one of the CP points (right to the IP), see 
comments in the text.}
\label{fig21}
\end{figure}

\begin{figure}[thb]
\centering
\includegraphics[width=2.5in,angle=0,trim=100 30 100 50]{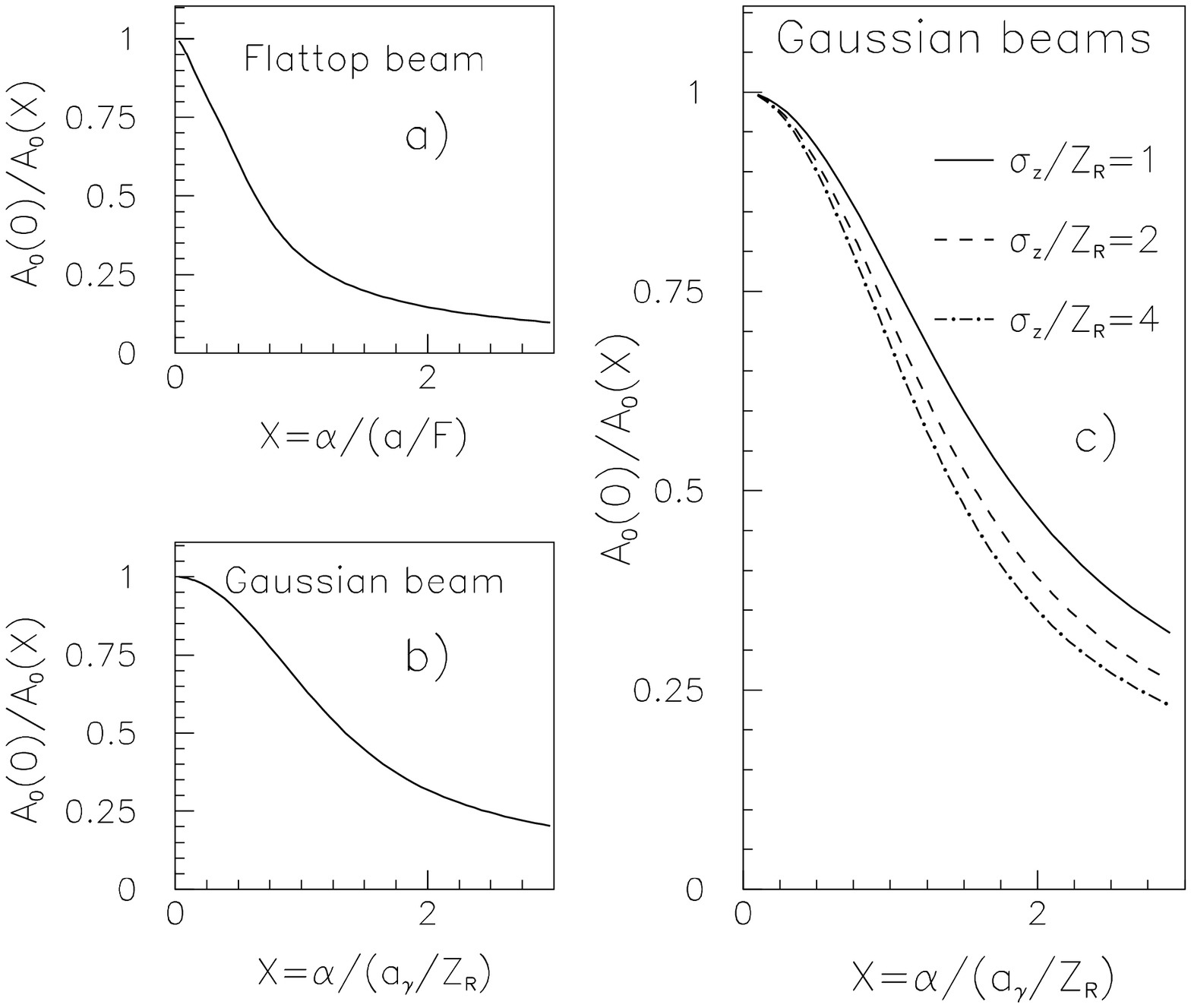}
\caption{Dependence of the required laser flash energy on 
the collision angle between the laser and electron beams;
(a) flattop, (b) Gaussian beams.
Plots a), b) correspond to very long laser beams. Plot c) shows the same 
for Gaussian beams for different ratios of $\sigma_z/Z_R$ assuming
that $\sigma_{L,z}=\sigma_z$.}
\label{fig22}
\end{figure}

In this section it is  considered how to bring laser beam into and out of
CP. Our solution is shown in Fig.~\ref{fig21}. What is essential here.

1) It is attractive to place the focusing mirror out of the incoming and 
outgoing electron beams. Unfortunately in this case the required
laser flash energy should be larger by a factor of 3 (the dependence
of the flash energy on the angle between the laser and electron beam is 
shown in fig.~\ref{fig22} for flattop and Gaussian beams). Therefore
we have chosen the head-on collision. In this case each of focusing
mirrors have two holes for the incoming and outgoing beams.

2) The opening angles of the final mirror M2 is dictated by the optimum
laser divergence at the CP. The calculations are performed in 
section 3. The minimum distance between the M2 mirror and IP 
is determined by the mirror damage threshold. For $A=3$~J, 
$\alpha_{\gamma, x}=0.018$ (for $Z_R=0.25$~mm, $\lambda=1\;\mu$m)
it would be sufficient to take this distance $\sim85$~cm.
In this case the fluence is 0.2 J/cm$^2$, while the damage threshold 
is in the 0.7--2 J/cm range~\cite{NLC,STUART}.
However, we put the focusing mirror M2 at a much larger distance (180 cm)
to provide near normal incidence of the incoming laser beam (for
smaller aberrations). Laser pulse comes to M2 after
a second slightly defocusing
mirror M1 placed between the IP and M2. In the case of using a plate
mirror at the M1 location the larger opening angle of the shielding
mask is required. The diameter of the focusing mirror M2 is about
20 cm ($\sim3\alpha_{\gamma,x}$) and that of M1 is twice smaller.

3) Due to the crab crossing angle $\alpha_c=30$~mrad (see Fig.7) the
horizontal size of focusing mirrors is somewhat larger than the vertical one
to provide the exit pass of the opposing laser beam. As soon as
the distance between two CP points is small, the incoming and
outgoing laser beams are practically parallel, but somewhat shifted.
     
 The scheme shown in Fig.~\ref{fig21} describes the case of head-on
collisions of the laser and electron beams. Due to crab crossing
collisions of the electron beam it is advantageous to
have crab crossing of the laser-electron beam collision too
(it was explained in sec.3). Fig.~\ref{fig21} is valid for this
option after some shift of the mirrors.

\section{Lasers}

From section 3 follows that for obtaining the conversion probability
$k\sim65$\% at $x=4.8$ and $E_0=250$~GeV a laser with the following 
parameters is required:

\begin{center}
\begin{tabular}{ll}
Power & $P\sim0.7$~TW \\
Duration & $\tau(rms)\sim\sigma_z/c\sim 1 - 2.5$~ps \\
Flash energy & 2 - 4 J \\
Repetition rate & collision rate at a collider \\
Average power & $\sim 25$~kW \\
Wave length & $\lambda=4.2E_0$[TeV], $\mu$m
\end{tabular}
\end{center}

\noindent Obtaining of such parameters is possible with  solid
state or free electron lasers (FEL). For $\lambda\ge1\mu$m
($E_0\gtrsim250$~GeV) FEL is the only option seen now.

\subsection{Solid state lasers}

The region of $\lambda\sim1\;\mu$m is convenient for solid state lasers,
namely this is a wave length  of the most powerful Nd:Glass lasers. 
In the last 10 years the technique of short powerful lasers
made an impressive step and has reached petowatt ($10^{15}$)
power levels and few femtosecond durations~\cite{PERRY}.
Obtaining  few joule pulses of picosecond duration is not
a problem for a modern laser technique. For photon collider
applications the main problem is a high repetition rate. 
This is connected with overheating of the media.

  The success in obtaining picosecond pulses is connected with a chirped 
pulse amplification (CPA) technique~\cite{STRIC}. ``Chirped'' means
that the pulse has a time-frequency correlation. The main problem in 
obtaining short pulses is the limitation on peak power imposed
by the nonlinear refractive index. This limit on intensity is about
1 GW/cm$^2$. CPA technique succesfully overcomes this limit.

\begin{figure}[!hbt]
\centering
\includegraphics[width=5.3in,angle=0,trim=0 0 0 0]{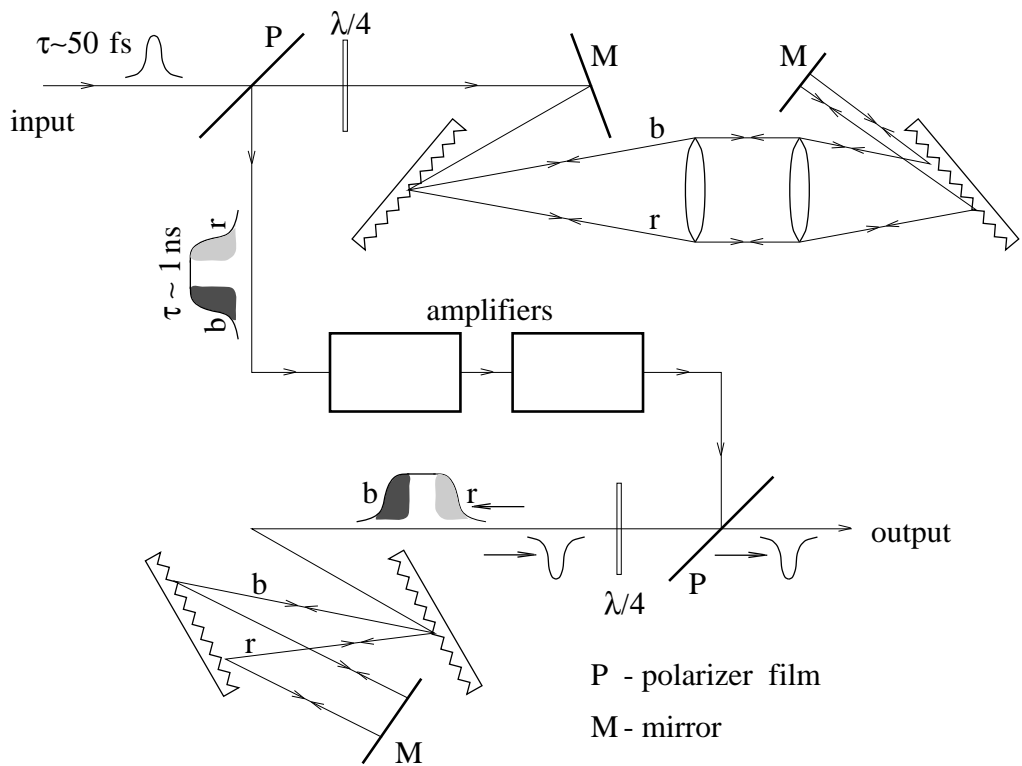}
\caption{Chirped pulse amplification. }
\label{fig23}
\end{figure}

 The principle of CPA is demonstrated in Fig.~\ref{fig23}.
A short, low energy pulse is generated in an oscillator.
Then  this pulse is stretched by a factor about $10^4$ in the grating
pair which has delay proportional to the frequency. This long
 nanosecond pulse is amplified and compressed by another grating pair 
to a pulse with the  initial or somewhat longer duration. Due to practical 
absence of non-linear effects, the obtained pulses have a very good 
quality close to the  diffraction limit.

   One of such lasers~\cite{BAMBER} works now in the E-144 experiment
studing nonlinear QED effects in the collision of laser photons
and 50 GeV electrons. It has a repetition rate of 0.5 Hz, 
$\lambda=1.06\;\mu$m (Nd:Glass), 2J flash energy, 2 TW power and
1 ps duration. This is a top-table laser. Its parameters are 
very close to our needs, only the repetition rate is too low.

   In this laser a flashlamp pumping is used. Further progress 
in the repetition rate (by two orders) is possible with a diode 
pumping (high efficiency semiconductor lasers). This technology is
fast developing and promoted by other big projects. 
With diode pumping the efficiency of solid state lasers reaches
a 10\% level. Recent studies~\cite{MEYER95,CLAY95,NLC} have shown
that the combination of CPA, diode pumping, recombining of several
lasers (using polarizers and Pockel cells or slightly different
wave lengths), and (if necessary) other laser techniques such 
as phase-conjugated mirrors, moving amplifiers allows already now
to build a solid state laser system for a photon collider.  
  
   All necessary technologies are developing actively now
for other (than HEP) applications. The detailed design of the solid
state laser system for the \GG, \GE\ colliders requires special
R\&D study. 
  

\subsection{Free Electron Lasers}
\label{sec:fel-parameters}

   Potential features of a free electron laser (FEL) allow one to 
consider it as an ideal source of primary photons for a gamma-gamma 
collider. Indeed, FEL radiation is tunable and has always minimal 
(i.e.  diffraction) dispersion.  The FEL radiation is completely 
polarized:  circularly or linearly for the case of helical or planar 
undulator, respectively.  A driving accelerator for the FEL may be a 
modification of the main linear accelerator, thus providing the 
required time structure of laser pulses.  The problem of 
synchronization of the laser and electron bunches at the conversion 
region is solved by means of traditional methods used in accelerator 
techniques. A FEL amplifier has potential to provide high conversion 
efficiency of the kinetic energy of the electron beam into coherent 
radiation. At sufficient peak power of the driving electron beam the 
peak power of the FEL radiation could reach the required TW level.

\begin{figure}[!htb]
\centering
\includegraphics[width=4in,angle=0]{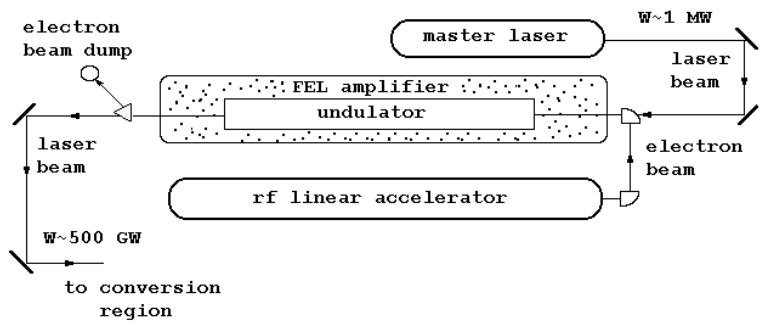}
\caption{MOPA FEL configuration for a gamma-gamma collider.} 
\label{fig24}
\end{figure}

   The idea to use a FEL as a laser for the gamma-gamma collider has been 
proposed in ref.~\cite{KONDR82}. A more detailed study of this idea has 
shown that the problem of construction of free electron laser can be 
solved using Master Oscillator -- Power Amplifier (MOPA) scheme with 
the driving accelerator for the FEL amplifier constructed on the same 
basis as the main accelerator for a linear collider 
\cite{fvp13,fvp13-1,tev-plc,slc-plc-nima}. At present an option of  
FEL as a laser for the gamma-gamma collider is studied for different 
projects. While there exist different FEL configurations, an 
amplifier configuration has definite advantages for application in the 
gamma-gamma collider schemes \cite{slc-plc-nima}.  The choice of 
specific technical solution depends on the parameters of the linear 
collider project. For instance, for the VLEPP, CLIC, JLC and TESLA 
projects it has been considered to use MOPA FEL scheme 
\cite{balakin-gg-1,tesla-plc-desy,SAL95,hiramatsu-gg,mikhailichenko-gg}.  
Designers of NLC project consider a FEL scheme using an induction linac 
and chirped pulse amplification technique \cite{kim-chirp,nlc-cdr}. 

\begin{figure}[!hb]
\centering
\includegraphics[width=4in,angle=0]{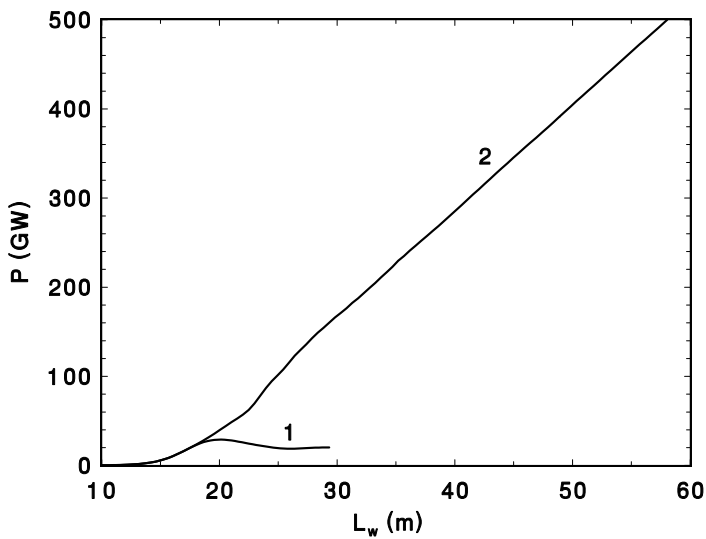}
\caption{
Output power of the FEL amplifier 
versus undulator length. 
} 
\label{fig25}
\end{figure}

   In the present study of the Linear Collider Project it has been 
accepted to use MOPA FEL scheme as a laser (see 
Fig.~\ref{fig24}).  Such a choice fits well to both TESLA and 
SBLC options. In this scheme the optical pulse from Nd glass laser 
($\lambda = 1 \ \mu$m, 1~MW peak power) is amplified by the FEL amplifier 
up to the power of about 500~GW (see Table~\ref{table4}). The 
driving beam for the FEL amplifier is produced by the linear rf 
accelerator identical to the main accelerator, but with lower 
accelerating gradient due to the higher beam load. It is important that the 
requirements to the parameters of the FEL driving electron beam are 
rather moderate and can be provided by an injector consisting of 
gridded thermoionic gun and subharmonic buncher \cite{holtkamp-rf-gun}. 

\begin{table}[!htb]
\caption{Parameters of the FEL amplifier}

\begin{tabular} { l p{.2cm} l }
\hline \\
\underline{Electron beam}$^*$ \\      
\hspace*{6pt} Electron energy                    &&  2 GeV \\      
\hspace*{6pt} Peak beam current                  &&  2.5 kA \\ 
\hspace*{6pt} rms energy spread                  &&  0.2 \% \\
\hspace*{6pt} rms normalized emittance           && $2\pi \times 10^{-2}$ 
                                                    cm rad \\
\hspace*{6pt} rms bunch length                   && 1~mm \\ 
\underline{Undulator} \\
\hspace*{6pt} Undulator type                     &&  Helical \\
\hspace*{6pt} Undulator period (entr.)           &&  15 cm  \\
\hspace*{6pt} Undulator field (entr.)            &&  10.2 kG \\
\hspace*{6pt} External $\beta$-function          &&  2 m \\
\hspace*{6pt} Length of untapered section        && 17 m \\ 
\hspace*{6pt} Total undulator length             && 60 m \\
\underline{Radiation} \\
\hspace*{6pt} Radiation wavelength               &&  1 $\mu $m \\
\hspace*{6pt} Input power                        &&  1 MW  \\
\hspace*{6pt} Output power                       &&  500 GW \\
\hspace*{6pt} Flash energy                       &&  2.3~J \\
\hspace*{6pt} Efficiency                         &&  10 \% \\
\hline \\ 
\end{tabular} 

$^*$Time diagram of the accelerator operation is identical to the 
time diagram of the main accelerator. 
\label{table4}
\end{table}

    Table~\ref{table4} presents the main parameters of the FEL 
amplifier and the driving accelerator. It is seen from 
Fig.~\ref{fig25} that 500~GW level of output power is achieved at 
the undulator length of about 60~m. The total flash energy in the laser 
pulse is about of 2~J.  

    Using the FEL amplifier allows one to completely exclude transmitting 
optical elements and deliver the laser beam to the conversion region 
using several reflections from metallic mirrors which are rather stable 
to the laser radiation damage. This can be done when vacuum systems of 
the FEL amplifier and linear collider are combined. The first 
reflection mirror can be installed at a distance about several tens of 
meters after the exit of the undulator when the laser beam expands 
to the size about several centimeters. 

   To reduce the cost of laser system, only one free electron laser can 
be used. This scheme operates as follows.  The FEL is installed only in 
one branch of the linear collider. When the laser bunch passes the 
focus of the conversion region, it is not dumped but is directed to the 
optical delay line which provides a delay time equal to the time 
interval between the bunches. Then it is focused on the electron beam 
of the opposite branch of the linear collider. Of course, this 
configuration provides colliding gamma-beams with the second micropulse 
of the collider.  Nevertheless, the number of microbunches is equal to 
several hundreds, so it will not result in significant reduction of the 
integral luminosity.  

\subsubsection{Future perspectives}

\begin{figure}[!hbt]
\centering
\includegraphics[width=4in,angle=0]{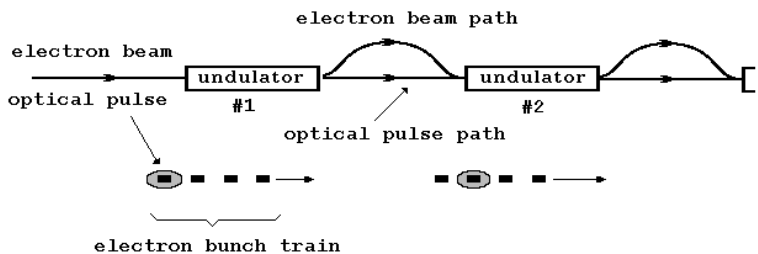}
\caption{The scheme of multi-stage FEL amplifier. One optical pulse is 
amplified by a sequence of electron bunches. The peak power of the 
output radiation exceeds by a factor of $N$ (number of 
amplification stages) the peak output radiation power of traditional 
single pass FEL amplifier.
} 
\label{fig26}
\end{figure}

    The present design has been limited with an approach which can be 
realized at the present level of accelerator and FEL technique. The 
main reserve to improve the FEL performance is to increase its 
efficiency which will allow to decrease the requirements to the value 
of the peak electron beam power. The perspectives of the FEL 
efficiency increase are on the way of using multi-stage FEL amplifier 
with diaphragm focusing line (see Fig.~\ref{fig26}) 
\cite{icf,SAL95}. The principle of operation of this FEL scheme consists 
in the storing of the energy in a single laser pulse amplified by a 
sequence of electron bunches. This scheme has evident perspectives for 
the TESLA project due to a large bunch spacing. Preliminary study shows 
that the energy of the driving electron beam could be reduced to the 
value of several hundreds of MeV and the value of the peak current 
could be reduced by several times.  The beam load in accelerator will 
be also reduced approximately by a factor of 3 due to higher FEL 
efficiency about of 30~\%. 

\end{document}